\documentclass[12pt,preprint]{aastex}

\begin{document}

\title{On the Late-Time Spectral Softening Found in X-ray Afterglows of Gamma-Ray Bursts}

\author{\sc Yuan-Zhu Wang\altaffilmark{1,2}, Yinan Zhao\altaffilmark{3}, Lang Shao\altaffilmark{4,5}, En-Wei Liang\altaffilmark{1,2}, Zu-Jia Lu\altaffilmark{1,2}}
\altaffiltext{1}{GXU-NAOC Center for Astrophysics and Space Sciences, Department of Physics, Guangxi University, Nanning 530004,
China}
\altaffiltext{2}{Guangxi Key Laboratory for Relativistic Astrophysics, Nanning, Guangxi 530004, China}
\altaffiltext{3}{Department of Astronomy, University of Florida, Gainesville FL 32611, USA}
\altaffiltext{4}{Department of Space Sciences and Astronomy, Hebei Normal University, Shijiazhuang 050024, China; lshao@hebtu.edu.cn}
\altaffiltext{5}{Key Laboratory of Dark Matter and Space Astronomy, Purple Mountain Observatory, Chinese Academy of Sciences, Nanjing 210008, China}

\begin{abstract}
Strong spectral softening has been revealed in the late X-ray afterglows of some gamma-ray bursts (GRBs). The scenario of X-ray scattering around circum-burst dusty medium has been supported by previous works due to its overall successful prediction of both the temporal and spectral evolution of some X-ray afterglows. To further investigate the observed feature of spectral softening, we now systematically search the X-ray afterglows detected by X-Ray Telescope (XRT) of {\it Swift} and collect twelve GRBs with significant late-time spectral softening. We find that dust scattering could be the dominant radiative mechanism for these X-ray afterglows regarding their temporal and spectral features. For some well observed bursts with high-quality data, their time-resolved spectra could be well produced within the scattering scenario by taking into account the X-ray absorption from circum-burst medium. We also find that during spectral softening the power-law index in the high energy end of the spectra does not vary much. The spectral softening is mainly manifested by the spectral peak energy continually moving to the soft end.

\end{abstract}
\maketitle

\section{Introduction}

Thanks to the {\it Swift} satellite (Gehrels et al. 2004), which has been in service for over ten years, our knowledge on gamma-ray bursts (GRBs) has been greatly extended especially in the X-ray wavelength. The overall light curves of X-ray afterglows have been revealed to be somewhat puzzling with diverse physical origins (e.g., Zhang et al. 2006; Nousek et al. 2006; Liang et al. 2007) especially when the optical afterglow is also taken into account (Panaitescu et al. 2006; Liang et al. 2008). In general, the observed multi-wavelength afterglows have been found consistent with the external forward shock models (e.g., Piran 2004; M\'{e}sz\'{a}ros 2006 for reviews). In contrast to the remarkable variations of the X-ray light curves, most spectra of X-ray afterglows show little variation (Butler \& Kocevski 2007; Evans et al. 2009; Shao et al. 2010) which is consistent with the prediction of standard external forward shock models.

The first explicitly reported spectral variation was in the X-ray afterglow of an unusual X-ray flash XRF~060218 (e.g., Soderberg et al. 2006; Fan et al. 2006; Butler 2007). Later, an optically dark burst GRB~090417B showed significant softening after $\sim 2\times10^4$~s since the burst trigger (Holland et al. 2010). Very recently, a similar spectral softening was also reported in GRB~130925A (Zhao \& Shao 2014; Evan et al. 2014). The X-ray afterglow of GRB~090417B and 130925A are both found to be consistent with the previously proposed X-ray scattering scenario regarding their light curves and spectra (Shao \& Dai 2007; Shao et al. 2008). As it has been pointed out by Evans et al. (2014), this spectral behavior have also been detected in several other bursts. In the literature, GRB~100316D also showed presence of very soft X-ray emission similar to XRF~060218 (Margutti et al. 2013).

In this paper, we collect a sample of twelve bursts well observed by the X-Ray Telescope (XRT) of {\it Swift} which showed a significant spectral softening at a late time ($\geq 10^4$~s) since the burst trigger. We will show that the radiative features regarding their light curves and spectral evolution are very consistent with the X-ray scattering scenario. We will make an effort to study their time-resolved spectra focusing on the radiative feature of this spectral softening. Our burst sample and data analysis is described in section 2. The X-ray light curves and spectral evolution of these bursts are analyzed with the scattering model in section 3. The time-resolved spectra are further analyzed and reproduced by the scattering model in section 4. Discussions and conclusion are given in section~5.

\section{Sample selection and data analysis}

To select the bursts that show significant spectral evolution in the X-ray afterglow, we made use of the Burst Analyser data\footnote{$http://www.swift.ac.uk/burst\_analyser/$} from UK {\it Swift} Science Data Centre (Evans et al. 2010). In order to study the spectral details of late-time afterglow, the bursts we choose should last long enough and be bright enough to make time-sliced spectra. To satisfy this, the sample was selected and handled as follows.

First, we checked the displayed results of Burst Analyzer on Swift UK site to find the evidence for late-time spectral evolution, i.e., the detectable changes of hardness ratios after $10^{4}$s (since when most of the early X-ray flares have faded away). As a result, we found 28 bursts to have this kind of behavior up to October 2013 which all showed hard-to-soft spectral evolution.

Then, we used the server of the Swift UK site to extract a series of time-sliced spectra after $10^4$~s (considered as late epoch in this work) for these 28 bursts. We adopted the scheme for slicing time bins introduced by Zhang et al. (2007). In most of the cases, we would want to have the spans of time intervals of each burst to be the equal in logarithmic scale. For instance, the first time interval starts at 10000s and ends at 20000s. Then the spans of the following intervals form a geometric progression $\Delta T_i=2^{i-1}\Delta T_0$ for each burst, where $\Delta T_i$ is the span for time interval $i$ and $\Delta T_0=10000$~s. In order to perform reliable spectral fitting, the total counts in each time interval should be greater than 100. If the total counts in one time interval are less than 100, we combine the interval with the next one. We also extracted the time intervals of 100s $\sim$ 10000s (considered as early epoch) for each burst in the similar way.

The XSPEC ver.12.8.1 (Arnaud 1996) included in HEASOFT ver.6.14 was used to fit the spectrum of each time interval. We consider a simple power law model, combined with the absorption from the host galaxy and milky way respectively, i.e. PHABS*zPHABS*Powerlaw for the bursts with redshifts detected, and PHABS*PHABS*Powerlaw for the bursts whose redshifts are unknown. The first PHABS was fixed at the Galactic value for each burst. Considering the the intrinsic absorption might not be varying a lot during the appearance of a burst, the second PHABS was left free but constant within the same burst. The power-law index was free, as well as the normalization. All the spectra from the same burst and same epoch (early or late) were fitted simultaneously using W-statistic in the XSPEC. As suggested in the Appendix B of XSPEC manual\footnote{$https://heasarc.gsfc.nasa.gov/xanadu/xspec/manual/XSappendixStatistics.html$}, W-statistic might generate uncalled-for wrong best fit for some weak sources and binning to ensure that every bin contains at least one count would often fix the problem. As a conservative approach, we rebinned all of the data using grppha in Science Tools to $>$ 20 counts per bin for early-epoch spectra and to $>$ 5 counts per bin for late-epoch spectra, respectively (some burst with low counts rate were rebinned to $>$ 5 and $>$ 2 counts per bin for early and late epoches, respectively). The systematic uncertainty that might be introduced by this rebinning scheme is still uncertain and the best-fit results should be taken with caution.

Our aim for this data analysis is to collect the bursts with significant late-time spectral softening after $10^{4}$~s. We judge the presence of significant spectral softening by comparing the first and the last spectral indices derived from the preliminary spectral fitting. The softening is considered significant if the 90 percent confidence intervals for the two spectral indices do not overlap with each other and the later turn out to be softer than the former. As a result we found 12 bursts that could meet these criteria and obtained 111 spectra in total as listed in Table~1. Their light curves and the spectral power-law indices in different time intervals are shown in Figure~1.

\section{Modeling Light curves and spectral evolution}

The GRB afterglows are generally considered as being radiated by the relativistic electrons accelerated in the external shocks due to relativistic GRB ejecta propagating in circum-burst medium (e.g., Piran 2004; M\'{e}sz\'{a}ros 2006, for reviews). To successfully interpret the seemingly very complicated afterglow light curves, it would take great effort to develop the external shock models (e.g., Li et al. 2015; Wang et al. 2015). Alternatively, for the X-ray afterglows that have shallow decay in the light curves and softening in the spectra, an X-ray scattering scenario (Shao \& Dai 2007) has been proposed to nicely reproduce both the light curves and spectral evolution (Shao et al. 2008; Holland et al. 2010; Zhao \& Shao 2014; Evans et al. 2014). In the scenario, a severe optical extinction would also be predicted (Shen et al. 2009).

In the previous works, to better fit the light curves of GRB~090417B and 130925 (Holland et al. 2010; Zhao \& Shao 2014), a smaller size upper limit ($a_+\sim 0.3\,{\rm \mu m}$) of the dust grains typically found in interstellar medium (ISM) would be suggested. But to be consistent with the evolution of spectral indices, a relatively harder initial spectral index of the prompt emission in X-ray wavelength would be required, which may indicate the self-absorbing processes taking place in the prompt emission (Holland et al 2010). Here we would further investigate these physical parameters by fitting the light curves and spectral evolution of our extended sample with the dust scattering model.

We adopt the algorithm introduced in Zhao \& Shao (2014) to calculate the radiative flux of scattered X-ray photons off circum-burst dust grains, which looses the Reyleigh-Gans (RG) limit to allow dust grains with larger size to be involved. To compare the spectral evolution predicted by the model with the observational data, in the literature, it is straightforward to compute the photon index predicted by the model and compare that with the observational one (Shen et al. 2009; Holland et al. 2010; Zhao \& Shao 2014). For simplicity, we adopt the ``pseudo'' spectral index as introduced in Shen et al. (2009). The ``pseudo'' spectral index $\Gamma$ is determined by fitting a power-law only with the two flux densities at the two ends of the spectrum, say, at 0.3 and 10 keV, respectively, for {\it Swift}/XRT.

In our convention, the spectral shape of prompt X-ray emission from GRBs has the form of $S(E)\propto E^ \delta$ as in Eq.~(9) of Zhao \& Shao (2014). The scattered X-ray emission that we receive in the detector would have the spectral shape, i.e., flux density $F_E(t)$ at a given time $t$ described by Eq.~(8) in Zhao \& Shao (2014). For completeness, we write down the form of $F_E(t)$ as
\begin{eqnarray}
F_{E}(t)=\int_{a_-}^{a_+}
S(E)\frac{{\rm d}N}{{\rm d}a}\frac{c\pi a^2}{R}\left(\frac{2\pi E a}{hc}\right)^2\left|A(\hat{\rho}, \hat{\theta})\right|^2{\rm d}a.
\end{eqnarray}
The complex amplitude function $A(\hat{\rho}, \hat{\theta})$ has been introduced by van de Hulst (1957) and further addressed in Zhao \& Shao (2014). For completeness, we rewrite its form here with a little rearrangement of the symbols as the following
\begin{eqnarray}
A(\hat{\rho},\hat{\theta})=\int_{0}^{\frac{\pi}{2}}\left(1-{\rm e}^{-i\hat{\rho}\sin\tau}\right)J_{0}\left(\hat{\theta}\cos\tau\right)
\cos\tau \sin\tau{\rm d}\tau,
\label{eq:ca}
\end{eqnarray}
where the item $\hat{\rho}$ is the phase shift of the photon with energy $E$ in the dust grain with a size $a$ as given by
\begin{eqnarray}
\hat{\rho}\simeq 3\times \left(\frac{1+z}{2}\right)^{-1}  \left(\frac{E}{1\,{\rm keV}}\right)^{-1} \left(\frac{a}{1\,{\rm \mu m}}\right),
\label{eq:rho}
\end{eqnarray}
and the item $\hat{\theta}$ is the dimensionless scattering angle for dust grains located at the distance $R$ from the GRB source as given by
\begin{eqnarray}
\hat{\theta}=\frac{2\pi Ea}{hc}\sqrt{\frac{2(1+z)ct}{R}},
\label{eq:sa}
\end{eqnarray}
where $h$ is the Planck constant and $c$ is the speed of light in the vacuum. The light curve would then be determined by the integral $F(t)=\int F_E(t) dE$ in a given wavelength range, e.g., 0.3-10 keV for XRT.

As shown in Figure~1, most of the bursts\footnote{See Zhao \& Shao (2014) for the plot of GRB~130925.} have a significant hard-to-soft evolution almost right after $\sim 10^4$~s.  Many of them even become very soft with a change in spectral index of $\Delta \Gamma >2$. These features at late time violate the prediction of standard external shock models, which consider the afterglow as the synchrotron radiation by relativistic electrons accelerated in external shocks. The simultaneous fitting for the time intervals before $10^{4}$s was not successful. Most of earlier X-ray emission may come from the early ``steep decay'' phase of the prompt emission with strong spectra evolution due to curvature effect (Zhang et al. 2007) or from the abnormal early X-ray flares (Chincarini et al. 2010).

As in Zhao \& Shao (2014), we fit the light curves and spectral evolution simultaneously with the dust scattering model for the data after $\sim10^4\,{\rm s}$. The best-fit model parameters are given in Table~2. Some physical parameters that would not change considerably during the fitting are given fixed values: $a_-=0.005\,{\rm \mu m}$ and $\beta=-3.5$. For the afterglows without known redshifts, we assume that $z=1$. As an interesting result, while the location of the dusty shell appear to be quite different for these bursts, the characteristic sizes of dust grains turn out to be typical ($\sim0.3\,{\rm \mu m}$) as in the ISM. The only burst that has significantly larger dust grains is GRB/XRF~060218 which is a low luminosity burst and had an association with a type-Ic supernova (e.g., Pian et al. 2006). However, there appears to have a degeneracy between the model parameters, especial between the location of the dusty shell $R$ and the maximum radius of dust grains $a_+$, as has been pointed out by Irwin \& Chevalier (2015). They proposed that a typical Galactic distribution of dust grain would also give a reasonably good fit to the data of GRB/XRF~060218 even though they made a small modification to the model by assuming a different source spectrum of the prompt emission.

Though most of the light curves can be well consistent with the models, we can see that some evolution of the spectral indices are not well reproduced. There might be a couple of issues that need to be mentioned. The first one is the difficulty in calculating these model light curves and spectral indices since multiple integrals over a series expansion are involved (Zhao \& Shao 2014). In this work, we have only obtained the maximum likelihood for model parameters by searching for the minimum chi-square in a manually chosen and evenly-sampled parameter space, instead of using a more sophisticated fitting scheme such as a Markov chain Monte Carlo (MCMC) method. Given that the model parameters also appear to have a degeneracy and the parameter space cannot be fully explored due to the fact that the process of the model evaluating is unavoidably timing-consuming, a true best fitting might be missed for some bursts. Therefore, what we have obtained here for these best-fit model parameters should be taken with a caution in their use. Nevertheless, the simultaneous fitting to both the light curve and spectral evolution has been very promising and the resulting physical parameters might have shown valuable information about the circum-burst medium. Further studies over the grain size and location of the dusty shells would provide much information on the GRB progenitors (e.g., Weingartner \& Draine 2001; Ramirez-Ruiz et al. 2001). As suggested by the results of Zhao \& Shao (2014), if the size distribution of dust grains around GRB can be confirmed to be as typical as in the ISM, the evaluation scheme of the original model adopting the RG approximation (Shao \& Dai 2007; Shao et al. 2008) would be much simpler, and a more powerful fitting scheme such as MCMC method would be helpful to constrain the model parameters.

\section{Modeling time-resolved spectra}

Another reason for the evolution of the spectral indices not being well reproduced for some bursts might be that we have adopted only one simple ``pseudo'' spectral index in a narrow wavelength range (0.3 - 10 keV) for evaluating the spectral evolution as introduced above. Usually, the shape of a spectrum would be determined by more than one spectral parameters. Thanks to the publicity of {\it Swift} data, we could now be able to acquire the time-resolved spectral data of X-ray afterglows and compare them with our model predictions in greater detail. According to the original X-ray scattering scenario (Shao \& Dai 2007), the spectra of received X-ray emission caused by dust scattering should have a form as given by Eq.~(1). However, that form has not taken into account the X-ray absorption from the circum-burst medium, which might be a great factor shaping the spectra in the soft X-rays. We have known that the absorption from dust grains in X-rays could be usually neglected (e.g., Laor \& Draine 1993; L\u{u} et al. 2011). But the circum-burst gases are severe absorber to X-rays.
Therefore, to confront the model with the real observational data especially spectrum-wise, we need to revise our evaluation of the spectra by taking into account the photoelectric absorption of (mostly soft) X-ray photon by the gases in the circum-burst medium and/or ISM. This absorption effect has been extensively studied and already standardized in astrophysical softwares, such as in XSPEC (Arnaud 1996).

Given that we still have difficulties in transplanting the X-ray scattering model into XSPEC as a user-defined model, we have compromised in this work to adopt an analytical approximation for the effect of X-ray absorption to proceed our calculation. The optical depth due to photoelectric absorption in the ISM over different photon energy $E$ has the form of $\tau(E)=\sigma(E) N_{\rm H}$, where $\sigma(E)$ is the total photoionization cross section taking into account different ingredients including gases, molecules and dust grains in the ISM and $N_{\rm H}$ is the total neutral hydrogen column density in the ISM. In general, this could be a very complicated problem especially in the ISM around the vicinity of GRBs (e.g., Greiner et al. 2011; Littlejohns et al. 2015). For an approximate evaluation which makes our model self-consistent, we adopt the form
\begin{eqnarray}
\sigma(E)=\sigma_0\left[{E(1+z)\over 1\,{\rm keV}}\right]^{-\gamma},
\label{eq:sigma}
\end{eqnarray}
where we have $\sigma_0\simeq 10^{-21.5}\,{\rm cm}^2$ and $\gamma\simeq2.5$ as suggested by Wilms et al. (2000) for the accumulative absorption by the ISM with a same chemical composition as in our galaxy, and $z$ is the redshift of the GRB source. Here only the total hydrogen column density $N_{\rm H}$ is taken as a free parameter when interpreting the absorption in observed spectra, which is very similar to selecting the parameter zPHABS for the redshifted photoelectric absorption when using XSPEC for evaluating the spectral indices as introduced above.

Since the redshift is a major parameter in determining the quantitative spectrum especially in a narrow energy range less than two orders of magnitude, we now only interpret the time-resolved spectra of the bursts with known redshifts. This leads to seven bursts in our sample: GRB~060218 ($z=0.0331$; Mirabal \& Halpern 2006), GRB~080207 ($z=2.0858$; Hjorth et al. 2012), GRB~081221 ($z=2.26$; Salvaterra et al. 2012), GRB~100621 ($z=0.542$; Milvang-Jensen et al. 2010), GRB~111209 ($z=0.677$; Vreeswijk et al. 2011), GRB~130907 ($z=1.238$; de Ugarte Postigo et al. 2013) and GRB~130925 ($z=0.347$: Vreeswijk et al. 2013; Sudilovsky et al. 2013). However, GRBs~080207 is not considered in our spectral fitting. There are too few photons in the late-time spectra which therefore have very low signal-to-noise (S/N) ratios and would provide useless information on the model parameters. For a more convincing comparison between our model and the observational data, we now focus on the six bursts as listed in Table~3 which all have well determined time-resolved late-time spectra.

The time-resolved spectra of the six bursts are shown in Figure~2. The data access and analysis has been introduced in the section~2. For each burst with a redshift $z$, all the spectra at different time $t$ were fitted simultaneously according to the following formula
\begin{eqnarray}
F_E(t, z; N_{\rm H}, a_+, \delta, R)\propto {\rm exp}\left[-\sigma(E)N_{\rm H}\right]\times \int_{a_-}^{a_+}
\frac{E^{2+\delta} a^{0.5}}{R}\left|A(\hat{\rho}, \hat{\theta})\right|^2{\rm d}a.
\end{eqnarray}
Here in this formula the photoionization cross section $\sigma(E)$ is given above by Eq.~(5). For each burst, an constant coefficient at the beginning of the right-hand side of Eq.~(6) is assumed and taken as a free parameter. Therefore $t$ and $z$ all have pre-determined values for each spectrum in given time interval. The minimum grain size $a_-$ is not important and set as $a_-=0.005\,{\rm \mu m}$. Therefore, together with the constant coefficient, the neutral hydrogen column density $N_{\rm H}$, the maximum grain size $a_+$, the initial spectral index of prompt emission in X-ray band $\delta$, and the dust distance $R$ are also taken as free parameters. The constant coefficient determines the absolute flux level of the spectra group and the other four parameters determine the relative flux level between each time-sliced spectra in the group of each burst.

The best fits provided by Eq.~(6) for the time-resolved spectra of these bursts are shown in Figure~2 by solid lines in different colors from top to bottom as in from early to late. The corresponding time intervals for these spectra and the best-fit model parameters are listed in Table~3. All the time-resolved spectra of these six bursts can be well reproduced by the scattering model. The best-fit parameters here are in general consistent with those introduced in Section~3 based on light curves and spectral indices except that the spectral power-law index $\delta$ is slightly larger (harder) in the case of time-resolved spectral fitting. Based on the best-fit parameters, we confirm that the size of circum-burst dust grains tends to be as small as in the typical ISM. The distance between the central source and dusty shell is about 100~pc which is typical for the swept-up wind bubble surrounding late massive star (Castor et al. 1975; Ramirez-Ruiz et al. 2001; Mirabal et al. 2003).

Meanwhile, by fitting the X-ray spectra especially for the softer part, we can also more directly obtain the value of $N_{\rm H}$ self-consistently within the model. GRB~100621 was reported to have an intrinsic host extinction $A_{\rm V}=3.6\,{\rm mag}$ and an X-ray absorbing column of $N_{\rm H}=0.65\times10^{22}\,{\rm cm}^{-2}$ (Greiner et al. 2013). We have a similar $N_{\rm H}=0.6\times10^{22}\,{\rm cm}^{-2}$ based on our model-dependent fitting simultaneously to all the time-resolved spectra as in Table~3. The X-ray absorbing column of the host galaxy of GRB 130907A has been estimated as $N_{\rm H}=(0.98 \pm 0.11)\times10^{22}\,{\rm cm}^{-2}$ (Veres et al. 2014), which is close to our value of $0.4\times10^{22}\,{\rm cm}^{-2}$. Both bursts have suffered from significant dust extinction based on their values of $A_{\rm V}$. We have not evaluated the value of $A_{\rm V}$ for each burst in this work since it is more complicated and would involve more theoretical work on the dust extinction. For a pioneering work on the effect of dust extinction on optical afterglows, see L\u{u} et al. (2011). Melandri et al. (2012) classified GRB 081221 as a ``dark'' burst according to the slope of the spectral energy distribution between the optical and the X-ray band. The light curve of GRB 111209 is dominated by prompt, high-latitude and flaring emission until around $10^{5}\,{\rm s}$ after the trigger. The spectra can be fitted by the scattering model if we only focus on the X-ray afterglow after $10^{5}\,{\rm s}$, which may indicate that there might be a long-lasting additional component before that.

\section{Conclusion}

In this paper, we find that the late-time X-ray afterglows of the bursts in our sample appear to be overall consistent temporally and spectrally with the scattering scenario where the observed late-time X-ray emission comes from the scattering of early prompt X-ray emission off the circum-burst dust grains. The information on the circum-burst dusty medium can be determined by fitting the light curves and evolution of spectral indices with the scattering model first proposed by Shao \& Dai (2007) and further improved by Zhao \& Shao (2014). We have not tried to constrain the model parameters with sophisticated fitting scheme such as a MCMC metheod since the evaluation of the scattered emission is relatively time-consuming and not appropriate for the MCMC method. Our best fitting results indicate that almost all the bursts in our sample have a relatively small size distribution of dust grains as typical as in the ISM. This result is a little confusing since the grain size has been expected to be larger in the denser medium around GRBs(e.g., Weingartner \& Draine 2001). Although, our results also indicate that the distance of the dusty shells is very close to the dense wind bubble around late massive stars, say, a carbon-rich Wolf-Rayet star (e.g., Marston 1997; Chu et al.1999).

The major features predicted by the dust scattering model are the X-ray spectral softening and significant dust extinction in the optical (Shen et al. 2009). In our sample, all the GRBs have significant late-time spectral softening which is consistent with the first prediction. Besides, most of them tend to have indications of extra dust extinction in case the optical observation has been carried out which appears to be consistent with the second prediction (e.g., Evan et al. 2014). We have shown that the X-ray afterglows of these bursts in our sample are very consistent with one dominant radiative component. If some other radiative processes, such as the synchrotron radiation from the external shocks, exist in these bursts, they might be suppressed for some reason. The late-time spectral softening as in GRB~130925 were also proposed to be related with a blackbody component in addition to the typical power law spectrum (Piro et al. 2014). By time-resolved spectral analysis of the {\it Swift}/XRT data, we have not found any significant indication of a blackbody component at least before $\sim 10^6$~s of this burst. However, the last time interval of GRB~130925A after $\sim 10^6$~s did have a hardening spectral index (Zhao \& Shao 2014) which might need further inspection and requires some other explanation.

The significant late-time softening in the X-ray afterglows would have raised a great challenge to the external shock models. The light curves of most normal non-softening GRBs have been exclusively explained by the well-studied external shock models (e.g., Liang et al. 2007, 2008, 2009; Li et al. 2015; Wang et al. 2015). In principle, the dust scattering takes place at a distance of approximately a hundred parsecs, while the internal and external shocks would be produced at less than a parsec. Currently, we still have difficulties in having both the scattering model and shock model working together in one single burst event based on available observational data. Evans et al. (2014) has made an effort with a detailed discussion. The basic concern is that the circum-burst medium within one parsec would be relatively attenuate after being swept up by the massive stellar wind. The resulting circum-burst medium would be wind-like instead of uniform as typical as in the ISM. This would also raise an open issue for the circum-burst medium of GRBs, especially taking into account the unexpected ISM-like size distribution of dust grains as we have found in this paper. While this work was in preparation, Margutti et al. (2015) studied a sample of GRBs with soft ($\Gamma >3$) X-ray afterglow and identified a connection between the X-ray photon index $\Gamma$, the X-ray absorbing column density  $N_{\rm H}$ and the burst duration $T_{\rm 90}$. They proposed that the bursts with significant soft X-ray afterglows appeared to have significantly larger $N_{\rm H}$ and significantly longer prompt duration. This also raises an interesting concern to the radiative mechanism of the prompt emission.

In this work, we have made an effort to reproduce the time-resolved spectra of six bursts from {\it Swift}/XRT data with the dust scattering model without utilizing XSPEC or similar advanced softwares. To take into account the effect of X-ray absorption from the circum-burst medium, we have adopted a simple form for the total photonionization cross section as given in our Equation~(5), which is an analytical approximation to the numerical work by Wilm (2000). As we have shown above, we can nicely reproduce the shape of the time-resolved spectra in different time intervals within the dust scattering scenario assuming a constant hydrogen column density $N_{\rm H}$. It appears that, while the softening of the X-ray afterglow from dust scattering has been widely proposed in the literature, the high energy spectral index of the output spectrum does not vary much at all. E.g., this has been explicitly shown by Figure~4 of Shao \& Dai (2007). The spectral softening is mainly manifested by the spectral peak energy continually moving to the soft end. The X-ray absorption from the circum-burst medium may have played an import role in shaping the spectra. We will further investigate this issue in the following work.

\acknowledgments

We are grateful to the referee for helpful comments and suggestions to improve the presentation of this paper. This work made use of data supplied by the UK Swift Science Data Centre at the University of Leicester. This work was supported in part by the National Basic Research Program of China (No. 2014CB845800) and the National Natural Science Foundation of China (grants 11361140349 and 11103083).

\begin{deluxetable}{llllcccc}
\tabletypesize{\small}
\footnotesize
\tablewidth{0pc}
\tablecaption{Selected 12 GRBs with 111 time intervals for spectral analysis and the best-fit parameters of the single power-law model using XSPEC.}
\tablehead{\colhead{GRB} & \colhead{interval\tablenotemark{a}} & \colhead{epoch} & \colhead{$\Gamma$} &  \colhead{intrinsic $N_{\rm H}$\tablenotemark{b}} & \colhead{redshift\tablenotemark} & \colhead{w\_statistic/bins\tablenotemark} }
\startdata
060105	&	0.1	-	0.2	&	Early	&	$	2.08 	_{	-0.04 	}^{	+0.04 	}	$	&	$	0.36 	_{	-0.01 	}^{	+0.01	}	$	&	--	&	1164.58/1123 	\\
	&	0.2	-	0.4	&	-	&	$	1.96 	_{	-0.04 	}^{	+0.04 	}	$	&	$	-	_{		}^{		}	$	&	-	&	-	\\
	&	0.4	-	0.8	&	-	&	$	1.91 	_{	-0.04 	}^{	+0.04 	}	$	&	$	-	_{		}^{		}	$	&	-	&	-	\\
	&	0.8	-	1.4	&	-	&	$	1.87 	_{	-0.04 	}^{	+0.04 	}	$	&	$	-	_{		}^{		}	$	&	-	&	-	\\
	&	4.6	-	7.2	&	-	&	$	1.92 	_{	-0.08 	}^{	+0.08 	}	$	&	$	-	_{		}^{		}	$	&	-	&	-	\\
	&	10	-	20	&	Late	&	$	1.83 	_{	-0.12 	}^{	+0.12 	}	$	&	$	0.09 	_{	-0.04 	}^{	+0.04	}	$	&	-	&	301.81/372 	\\
	&	20	-	40	&	-	&	$	2.18 	_{	-0.18 	}^{	+0.18 	}	$	&	$	-	_{		}^{		}	$	&	-	&	-	\\
	&	40	-	80	&	-	&	$	2.12 	_{	-0.16 	}^{	+0.16 	}	$	&	$	-	_{		}^{		}	$	&	-	&	-	\\
	&	80	-	500	&	-	&	$	2.72 	_{	-0.23 	}^{	+0.22 	}	$	&	$	-	_{		}^{		}	$	&	-	&	-	\\
060218	&	0.1	-	0.2	&	Early	&	$	1.50 	_{	-0.07 	}^{	+0.07 	}	$	&	$	0.26 	_{	-0.01 	}^{	+0.01	}	$	&	0.033	&	4703.16/2355 	\\
	&	0.2	-	0.4	&	-	&	$	1.47 	_{	-0.03 	}^{	+0.03 	}	$	&	$	-	_{		}^{		}	$	&	-	&	-	\\
	&	0.4	-	0.8	&	-	&	$	1.40 	_{	-0.02 	}^{	+0.02 	}	$	&	$	-	_{		}^{		}	$	&	-	&	-	\\
	&	0.8	-	1.6	&	-	&	$	1.64 	_{	-0.01 	}^{	+0.01 	}	$	&	$	-	_{		}^{		}	$	&	-	&	-	\\
	&	1.6	-	2.8	&	-	&	$	2.28 	_{	-0.02 	}^{	+0.02 	}	$	&	$	-	_{		}^{		}	$	&	-	&	-	\\
	&	5.9	-	8.6	&	-	&	$	3.04 	_{	-0.12 	}^{	+0.12 	}	$	&	$	-	_{		}^{		}	$	&	-	&	-	\\
	&	10	-	20	&	Late	&	$	3.96 	_{	-0.38 	}^{	+0.43 	}	$	&	$	0.51 	_{	-0.08 	}^{	+0.10	}	$	&	-	&	293.99/265 	\\
	&	20	-	80	&	-	&	$	4.41 	_{	-0.39 	}^{	+0.46 	}	$	&	$	-	_{		}^{		}	$	&	-	&	-	\\
	&	80	-	1880&	-	&	$	5.34 	_{	-0.53 	}^{	+0.67 	}	$	&	$	-	_{		}^{		}	$	&	-	&	-	\\
080207	&	0.1	-	0.2	&	Early	&	$	1.31 	_{	-0.09 	}^{	+0.09 	}	$	&	$	1.06 	_{	-0.10 	}^{	+0.10	}	$	&	2.0858	&	325.45/327 	\\
	&	4.7	-	0.4	&	-	&	$	2.61 	_{	-0.20 	}^{	+0.20 	}	$	&	$	-	_{		}^{		}	$	&	-	&	-	\\
	&	10	-	0.8	&	Late	&	$	2.31 	_{	-0.24 	}^{	+0.22 	}	$	&	$	0.68 	_{	-0.12 	}^{	+0.12	}	$	&	-	&	142.70/122	\\
	&	20	-	1.4	&	-	&	$	3.03 	_{	-0.36 	}^{	+0.39 	}	$	&	$	-	_{		}^{		}	$	&	-	&	-	\\
081221	&	0.1	-	0.2	&	Early	&	$	2.23 	_{	-0.04 	}^{	+0.04 	}	$	&	$	3.32 	_{	-0.16 	}^{	+0.17	}	$	&	2.36	&	794.65/556 	\\
	&	0.2	-	0.4	&	-	&	$	2.31 	_{	-0.06 	}^{	+0.06 	}	$	&	$	-	_{		}^{		}	$	&	-	&	-	\\
	&	0.4	-	0.8	&		&	$	2.04 	_{	-0.05 	}^{	+0.05 	}	$	&	$	-	_{		}^{		}	$	&	-	&	-	\\
	&	4.8	-	6.6	&	-	&	$	1.94 	_{	-0.10 	}^{	+0.10 	}	$	&	$	-	_{		}^{		}	$	&	-	&	-	\\
	&	10	-	20	&	Late	&	$	2.22 	_{	-0.14 	}^{	+0.15 	}	$	&	$	4.97 	_{	-0.63 	}^{	+0.68	}	$	&	-	&	289.12/281 	\\
	&	20	-	40	&	-	&	$	2.36 	_{	-0.15 	}^{	+0.16 	}	$	&	$	-	_{		}^{		}	$	&	-	&	-	\\
	&	40	-	80	&	-	&	$	2.58 	_{	-0.31 	}^{	+0.32 	}	$	&	$	-	_{		}^{		}	$	&	-	&	-	\\
	&	80	-	600	&	-	&	$	3.49 	_{	-0.29 	}^{	+0.31 	}	$	&	$	-	_{		}^{		}	$	&	-	&	-	\\
090201	&	3.6	-	10	&	Early	&	$	2.00 	_{	-0.15 	}^{	+0.16 	}	$	&	$	0.47 	_{	-0.07 	}^{	+0.08	}	$	&	-	&	59.97/54 	\\
	&	10	-	20	&	Late	&	$	2.02 	_{	-0.14 	}^{	+0.15 	}	$	&	$	0.41 	_{	-0.04 	}^{	+0.05	}	$	&	-	&	447.78/407 	\\
	&	20	-	40	&	-	&	$	2.14 	_{	-0.15 	}^{	+0.15 	}	$	&	$	-	_{		}^{		}	$	&	-	&	-	\\
	&	40	-	80	&	-	&	$	2.39 	_{	-0.17 	}^{	+0.18 	}	$	&	$	-	_{		}^{		}	$	&	-	&	-	\\
	&	80	-	160	&	-	&	$	2.65 	_{	-0.24 	}^{	+0.25 	}	$	&	$	-	_{		}^{		}	$	&	-	&	-	\\
	&	160	-	320	&	-	&	$	2.95 	_{	-0.34 	}^{	+0.36 	}	$	&	$	-	_{		}^{		}	$	&	-	&	-	\\
	&	320	-	885	&	-	&	$	3.36 	_{	-0.47 	}^{	+0.50 	}	$	&	$	-	_{		}^{		}	$	&	-	&	-	\\
090404	&	0.1	-	0.2	&	Early	&	$	2.90 	_{	-0.07 	}^{	+0.07 	}	$	&	$	0.35 	_{	-0.02 	}^{	+0.02	}	$	&	-	&	652.57/347 	\\
	&	0.2	-	0.8	&	-	&	$	2.75 	_{	-0.30 	}^{	+0.30 	}	$	&	$	-	_{		}^{		}	$	&	-	&	-	\\
	&	0.8	-	1.25	&	-	&	$	2.69 	_{	-0.40 	}^{	+0.40 	}	$	&	$	-	_{		}^{		}	$	&	-	&	-	\\
	&	4.5	-	7.1	&	-	&	$	2.42 	_{	-0.16 	}^{	+0.16 	}	$	&	$	-	_{		}^{		}	$	&	-	&	-	\\
	&	10	-	20	&	Late	&	$	2.54 	_{	-0.18 	}^{	+0.19 	}	$	&	$	0.41 	_{	-0.05 	}^{	+0.05	}	$	&	-	&	290.26/280 	\\
	&	20	-	40	&	-	&	$	2.43 	_{	-0.20 	}^{	+0.20 	}	$	&	$	-	_{		}^{		}	$	&	-	&	-	\\
	&	40	-	80	&	-	&	$	2.94 	_{	-0.28 	}^{	+0.29 	}	$	&	$	-	_{		}^{		}	$	&	-	&	-	\\
	&	80	-	160	&	-	&	$	3.20 	_{	-0.30 	}^{	+0.31 	}	$	&	$	-	_{		}^{		}	$	&	-	&	-	\\
	&	160	-	320	&	-	&	$	3.67 	_{	-0.38 	}^{	+0.39 	}	$	&	$	-	_{		}^{		}	$	&	-	&	-	\\
	&	320	-	2000	&	-	&	$	4.05 	_{	-0.46 	}^{	+0.44 	}	$	&	$	-	_{		}^{		}	$	&	-	&	-	\\
100621	&	0.1	-	0.2	&	Early	&	$	2.75 	_{	-0.06 	}^{	+0.06 	}	$	&	$	1.74 	_{	-0.07 	}^{	+0.08	}	$	&	0.542	&	938.77/638 	\\
	&	0.2	-	0.4	&	-	&	$	2.53 	_{	-0.09 	}^{	+0.09 	}	$	&	$	-	_{		}^{		}	$	&	-	&	-	\\
	&	0.4	-	0.8	&	-	&	$	1.75 	_{	-0.18 	}^{	+0.19 	}	$	&	$	-	_{		}^{		}	$	&	-	&	-	\\
	&	0.8	-	2.35	&	-	&	$	2.06 	_{	-0.15 	}^{	+0.15 	}	$	&	$	-	_{		}^{		}	$	&	-	&	-	\\
	&	5.66	-	8.13	&	-	&	$	2.01 	_{	-0.15 	}^{	+0.15 	}	$	&	$	-	_{		}^{		}	$	&	-	&	-	\\
	&	10	-	20	&	Late	&	$	2.17 	_{	-0.16 	}^{	+0.16 	}	$	&	$	1.85 	_{	-0.18 	}^{	+0.19	}	$	&	-	&	453.51/457 	\\
	&	20	-	40	&	-	&	$	2.61 	_{	-0.16 	}^{	+0.17 	}	$	&	$	-	_{		}^{		}	$	&	-	&	-	\\
	&	40	-	80	&	-	&	$	2.59 	_{	-0.16 	}^{	+0.17 	}	$	&	$	-	_{		}^{		}	$	&	-	&	-	\\
	&	80	-	160	&	-	&	$	2.78 	_{	-0.25 	}^{	+0.26 	}	$	&	$	-	_{		}^{		}	$	&	-	&	-	\\
	&	160	-	320	&	-	&	$	3.26 	_{	-0.38 	}^{	+0.40 	}	$	&	$	-	_{		}^{		}	$	&	-	&	-	\\
	&	320	-	2000	&	-	&	$	3.38 	_{	-0.34 	}^{	+0.36 	}	$	&	$	-	_{		}^{		}	$	&	-	&	-	\\
110709	&	0.1	-	0.2	&	Early	&	$	2.23 	_{	-0.12 	}^{	+0.13 	}	$	&	$	0.74 	_{	-0.05 	}^{	+0.05	}	$	&	--	&	 587.44/543 	\\
	&	0.2	-	0.4	&	-	&	$	2.19 	_{	-0.11 	}^{	+0.11 	}	$	&	$	-	_{		}^{		}	$	&	-	&	-	\\
	&	0.4	-	0.635	&	-	&	$	2.04 	_{	-0.11 	}^{	+0.12 	}	$	&	$	-	_{		}^{		}	$	&	-	&	-	\\
	&	0.635	-	1	&	-	&	$	1.84 	_{	-0.17 	}^{	+0.17 	}	$	&	$	-	_{		}^{		}	$	&	-	&	-	\\
	&	1	-	2	&	-	&	$	1.93 	_{	-0.13 	}^{	+0.13 	}	$	&	$	-	_{		}^{		}	$	&	-	&	-	\\
	&	5.15	-	7.77	&	-	&	$	2.09 	_{	-0.12 	}^{	+0.12 	}	$	&	$	-	_{		}^{		}	$	&	-	&	-	\\
	&	10	-	20	&	Late	&	$	2.35 	_{	-0.22 	}^{	+0.23 	}	$	&	$	0.74 	_{	-0.12 	}^{	+0.13	}	$	&	-	&	114.64/125 	\\
	&	80	-	300	&	-	&	$	4.23 	_{	-0.62 	}^{	+0.69 	}	$	&	$	-	_{		}^{		}	$	&	-	&	-	\\
111209	&	0.425	-	0.8	&	Early	&	$	1.06 	_{	-0.02 	}^{	+0.02 	}	$	&	$	0.19 	_{	-0.01 	}^{	+0.01	}	$	&	0.677	&	2859.51/2514 	\\
	&	0.8	-	1.6	&	-	&	$	1.20 	_{	-0.01 	}^{	+0.01 	}	$	&	$	-	_{		}^{		}	$	&	-	&	-	\\
	&	1.6	-	2.07	&	-	&	$	1.12 	_{	-0.01 	}^{	+0.01 	}	$	&	$	-	_{		}^{		}	$	&	-	&	-	\\
	&	5.23	-	7.84	&	-	&	$	1.45 	_{	-0.01 	}^{	+0.01 	}	$	&	$	-	_{		}^{		}	$	&	-	&	-	\\
	&	10	-	20	&	Late	&	$	1.56 	_{	-0.02 	}^{	+0.02 	}	$	&	$	0.17 	_{	-0.01 	}^{	+0.01	}	$	&	-	&	1241.41/1300 	\\
	&	20	-	40	&	-	&	$	1.68 	_{	-0.05 	}^{	+0.05 	}	$	&	$	-	_{		}^{		}	$	&	-	&	-	\\
	&	40	-	80	&	-	&	$	2.00 	_{	-0.12 	}^{	+0.12 	}	$	&	$	-	_{		}^{		}	$	&	-	&	-	\\
	&	80	-	160	&	-	&	$	2.31 	_{	-0.12 	}^{	+0.12 	}	$	&	$	-	_{		}^{		}	$	&	-	&	-	\\
	&	160	-	320	&	-	&	$	2.46 	_{	-0.20 	}^{	+0.21 	}	$	&	$	-	_{		}^{		}	$	&	-	&	-	\\
	&	320	-	2560	&	-	&	$	2.75 	_{	-0.21 	}^{	+0.21 	}	$	&	$	-	_{		}^{		}	$	&	-	&	-	\\
120308	&	0.1	-	0.2	&	Early	&	$	2.80 	_{	-0.08 	}^{	+0.08 	}	$	&	$	0.10 	_{	-0.01 	}^{	+0.01	}	$	&	--	&	581.95/462 	\\
	&	0.2	-	0.285	&	-	&	$	3.89 	_{	-0.15 	}^{	+0.16 	}	$	&	$	-	_{		}^{		}	$	&	-	&	-	\\
	&	0.285	-	0.8	&	-	&	$	2.12 	_{	-0.15 	}^{	+0.15 	}	$	&	$	-	_{		}^{		}	$	&	-	&	-	\\
	&	0.8	-	1.6	&	-	&	$	1.75 	_{	-0.13 	}^{	+0.13 	}	$	&	$	-	_{		}^{		}	$	&	-	&	-	\\
	&	1.6	-	2.45	&	-	&	$	1.79 	_{	-0.13 	}^{	+0.13 	}	$	&	$	-	_{		}^{		}	$	&	-	&	-	\\
	&	5.84	-	10	&	-	&	$	1.76 	_{	-0.12 	}^{	+0.12 	}	$	&	$	-	_{		}^{		}	$	&	-	&	-	\\
	&	10	-	20	&	Late	&	$	1.66 	_{	-0.18 	}^{	+0.19 	}	$	&	$	0.08 	_{	-0.04 	}^{	+0.05	}	$	&	-	&	145.08/133 	\\
	&	20	-	40	&	-	&	$	1.93 	_{	-0.25 	}^{	+0.27 	}	$	&	$	-	_{		}^{		}	$	&	-	&	-	\\
	&	40	-	300	&	-	&	$	2.25 	_{	-0.38 	}^{	+0.40 	}	$	&	$	-	_{		}^{		}	$	&	-	&	-	\\
130907	&	0.1	-	0.2	&	Early	&	$	1.58 	_{	-0.04 	}^{	+0.04 	}	$	&	$	0.71 	_{	-0.02 	}^{	+0.02	}	$	&	1.238	&	2558.14/1845 	\\
	&	0.2	-	0.4	&	-	&	$	1.39 	_{	-0.02 	}^{	+0.02 	}	$	&	$	-	_{		}^{		}	$	&	-	&	-	\\
	&	0.4	-	0.8	&	-	&	$	1.59 	_{	-0.02 	}^{	+0.02 	}	$	&	$	-	_{		}^{		}	$	&	-	&	-	\\
	&	0.8	-	1.74	&	-	&	$	1.62 	_{	-0.01 	}^{	+0.01 	}	$	&	$	-	_{		}^{		}	$	&	-	&	-	\\
	&	7.63	-	7.89	&	-	&	$	1.72 	_{	-0.14 	}^{	+0.14 	}	$	&	$	-	_{		}^{		}	$	&	-	&	-	\\
	&	10	-	20	&	Late	&	$	1.72 	_{	-0.06 	}^{	+0.06 	}	$	&	$	0.71 	_{	-0.06 	}^{	+0.06	}	$	&	-	&	1099.01/1121 	\\
	&	20	-	40	&	-	&	$	1.79 	_{	-0.06 	}^{	+0.06 	}	$	&	$	-	_{		}^{		}	$	&	-	&	-	\\
	&	40	-	80	&	-	&	$	1.88 	_{	-0.07 	}^{	+0.07 	}	$	&	$	-	_{		}^{		}	$	&	-	&	-	\\
	&	80	-	160	&	-	&	$	2.08 	_{	-0.08 	}^{	+0.08 	}	$	&	$	-	_{		}^{		}	$	&	-	&	-	\\
	&	160	-	320	&	-	&	$	2.55 	_{	-0.20 	}^{	+0.21 	}	$	&	$	-	_{		}^{		}	$	&	-	&	-	\\
	&	320	-	640	&	-	&	$	2.66 	_{	-0.23 	}^{	+0.23 	}	$	&	$	-	_{		}^{		}	$	&	-	&	-	\\
	&	640	-	2560	&	-	&	$	3.45 	_{	-0.31 	}^{	+0.27 	}	$	&	$	-	_{		}^{		}	$	&	-	&	-	\\
130925	&	0.1	-	0.2	&	Early	&	$	1.89 	_{	-0.05 	}^{	+0.05 	}	$	&	$	1.73 	_{	-0.03 	}^{	+0.03	}	$	&	0.347	&	3773.76/2879 	\\
	&	0.2	-	0.4	&	-	&	$	1.79 	_{	-0.03 	}^{	+0.03 	}	$	&	$	-	_{		}^{		}	$	&	-	&	-	\\
	&	0.4	-	0.8	&	-	&	$	1.77 	_{	-0.03 	}^{	+0.03 	}	$	&	$	-	_{		}^{		}	$	&	-	&	-	\\
	&	0.8	-	1.5	&	-	&	$	1.70 	_{	-0.02 	}^{	+0.02 	}	$	&	$	-	_{		}^{		}	$	&	-	&	-	\\
	&	4.75	-	5.5	&	-	&	$	1.98 	_{	-0.02 	}^{	+0.02 	}	$	&	$	-	_{		}^{		}	$	&	-	&	-	\\
	&	6.68	-	7.27	&	-	&	$	1.59 	_{	-0.02 	}^{	+0.02 	}	$	&	$	-	_{		}^{		}	$	&	-	&	-	\\
	&	10	-	20	&	Late	&	$	2.53 	_{	-0.14 	}^{	+0.14 	}	$	&	$	2.07 	_{	-0.09 	}^{	+0.09	}	$	&	-	&	1200.25/1242 	\\
	&	20	-	40	&	-	&	$	2.99 	_{	-0.14 	}^{	+0.14 	}	$	&	$	-	_{		}^{		}	$	&	-	&	-	\\
	&	40	-	80	&	-	&	$	3.11 	_{	-0.11 	}^{	+0.11 	}	$	&	$	-	_{		}^{		}	$	&	-	&	-	\\
	&	80	-	160	&	-	&	$	3.46 	_{	-0.12 	}^{	+0.12 	}	$	&	$	-	_{		}^{		}	$	&	-	&	-	\\
	&	160	-	320	&	-	&	$	3.75 	_{	-0.12 	}^{	+0.12 	}	$	&	$	-	_{		}^{		}	$	&	-	&	-	\\
	&	320	-	640	&	-	&	$	3.99 	_{	-0.17 	}^{	+0.17 	}	$	&	$	-	_{		}^{		}	$	&	-	&	-	\\
	&	640	-	1280	&	-	&	$	4.33 	_{	-0.20 	}^{	+0.21 	}	$	&	$	-	_{		}^{		}	$	&	-	&	-	\\
	&	1280	-	6000	&	-	&	$	3.72 	_{	-0.21 	}^{	+0.21 	}	$	&	$	-	_{		}^{		}	$	&	-	&	-	\\

\enddata
\tablenotetext{a}{In the unit of $10^{3}$~s since the BAT trigger.}
\tablenotetext{b}{In the unit of $10^{22}\,{\rm cm}^{-2}$.}
\tablenotetext{-}{All the errors in this work indicate the 90 percent confidence intervals.}
\end{deluxetable}

\begin{deluxetable}{llllc}
\footnotesize
\tablewidth{0pc}
\tablecaption{The best-fit parameters of dust scattering model.}
\tablehead{\colhead{GRB} & \colhead{$a_+({\rm \mu m})$} & \colhead{$R({\rm pc})$} & \colhead{$\delta_1$\tablenotemark{a}} & \colhead{redshift} }
\startdata
060105  &  0.31 &	30  &	2.0  &	-- 	   	\\
060218	&  0.95 &	150	&  -1.0	 &	0.0331  \\
080207  &  0.33 &	20 	&   1.8  &  2.0858 	\\
081221  &  0.28 &	200	&   0.9  &	2.26   	\\
090201	&  0.29 &	100	&   1.4  &	-- 	   	\\
090404  &  0.33 &	300	&   0.5  &	-- 	   	\\
090417B\tablenotemark{b} &  0.25 &	30$\sim$80	&   2.0  &	0.345  	\\
100621  &  0.28 &	300	&   0.6  &	0.542  	\\
110709  &  0.31 &	50 	&   0.6  &	-- 	   	\\
111209  &  0.30 &	100	&   1.8  &	0.677  	\\
120308  &  0.31 &	50	&   1.9  &	--     	\\
130907  &  0.28 &	50 	&   1.9  &	1.238  	\\
130925\tablenotemark{c}  &  0.40 &	600	&  -0.3  &	0.347  	\\
\enddata
\tablenotetext{a}{Derived from the fitting over light curves and evolution of photon indices.}
\tablenotetext{b}{Results taken from Holland et al. (2010).}
\tablenotetext{c}{Results taken from Zhao \& Shao (2014).}
\end{deluxetable}

\begin{deluxetable}{llllccc}
\footnotesize
\tablewidth{0pc}
\tablecaption{The best-fit parameters for time-resolved spectral fitting.}
\tablehead{\colhead{GRB} & \colhead{interval\tablenotemark{a}} &\colhead{$a_+({\rm \mu m})$} & \colhead{$R({\rm pc})$} &  \colhead{redshift} & \colhead{$N_{\rm H}(10^{22}\,{\rm cm}^{-2})$} & \colhead{$\delta_2$\tablenotemark{b}} }
\startdata
060218  &   10  -   20      &  0.95 &	150	& 	0.0331 &    0.2  	&	0	\\
        &   20  -   80      &   --  &    -- &     --   &     --     &    -- \\
        &   80  -   1880    &   --  &    -- &     --   &     --     &    -- \\
081221  &   10  -   20      &  0.28 &	200	& 	2.26   &    2.5  	&	1.6	\\
        &   20  -   40      &   --  &    -- &     --   &     --     &    -- \\
        &   40  -   80      &   --  &    -- &     --   &     --     &    -- \\
        &   80  -   600     &   --  &    -- &     --   &     --     &    -- \\
100621  &   10  -   20      &  0.28 &	300	&  0.542   &    0.7  	&	1.5	\\
        &   20  -   40      &   --  &    -- &     --   &     --     &    -- \\
        &   40  -   80      &   --  &    -- &     --   &     --     &    -- \\
        &   80  -   160     &   --  &    -- &     --   &     --     &    -- \\
        &   160 -   320     &   --  &    -- &     --   &     --     &    -- \\
        &   320 -   2000    &   --  &    -- &     --   &     --     &    -- \\
111209  &   80  -   160     &  0.30 &	100	& 	0.677  &    0.01  	&	2.6	\\
        &   160 -   320     &   --  &    -- &     --   &     --     &    -- \\
        &   320 -   2560    &   --  &    -- &     --   &     --     &    -- \\
130907  &   10  -   20      &  0.28 &	50	&  1.238   &    0.4  	&	2.4	\\
        &   20  -   40      &   --  &    -- &     --   &     --     &    -- \\
        &   40  -   80      &   --  &    -- &     --   &     --     &    -- \\
        &   80  -   160     &   --  &    -- &     --   &     --     &    -- \\
        &   160 -   320     &   --  &    -- &     --   &     --     &    -- \\
        &   320 -   640     &   --  &    -- &     --   &     --     &    -- \\
        &   640 -   2560    &   --  &    -- &     --   &     --     &    -- \\
130925  &   20  -   40      &  0.30 & 1500	&  0.347   &    0.8  	&	1	\\
        &   40  -   80      &   --  &    -- &     --   &     --     &    -- \\
        &   80  -   160     &   --  &    -- &     --   &     --     &    -- \\
        &   160 -   320     &   --  &    -- &     --   &     --     &    -- \\
        &   320 -   640     &   --  &    -- &     --   &     --     &    -- \\
        &   640 -   1280    &   --  &    -- &     --   &     --     &    -- \\
\enddata
\tablenotetext{a}{Intervals for time-resolved spectral analysis with a unit of $10^{3}$s.}
\tablenotetext{b}{Derived from time-resolved spectral fitting.}
\end{deluxetable}

\begin{figure*}[htbp]
\centering

\includegraphics[angle=0,scale=0.35]{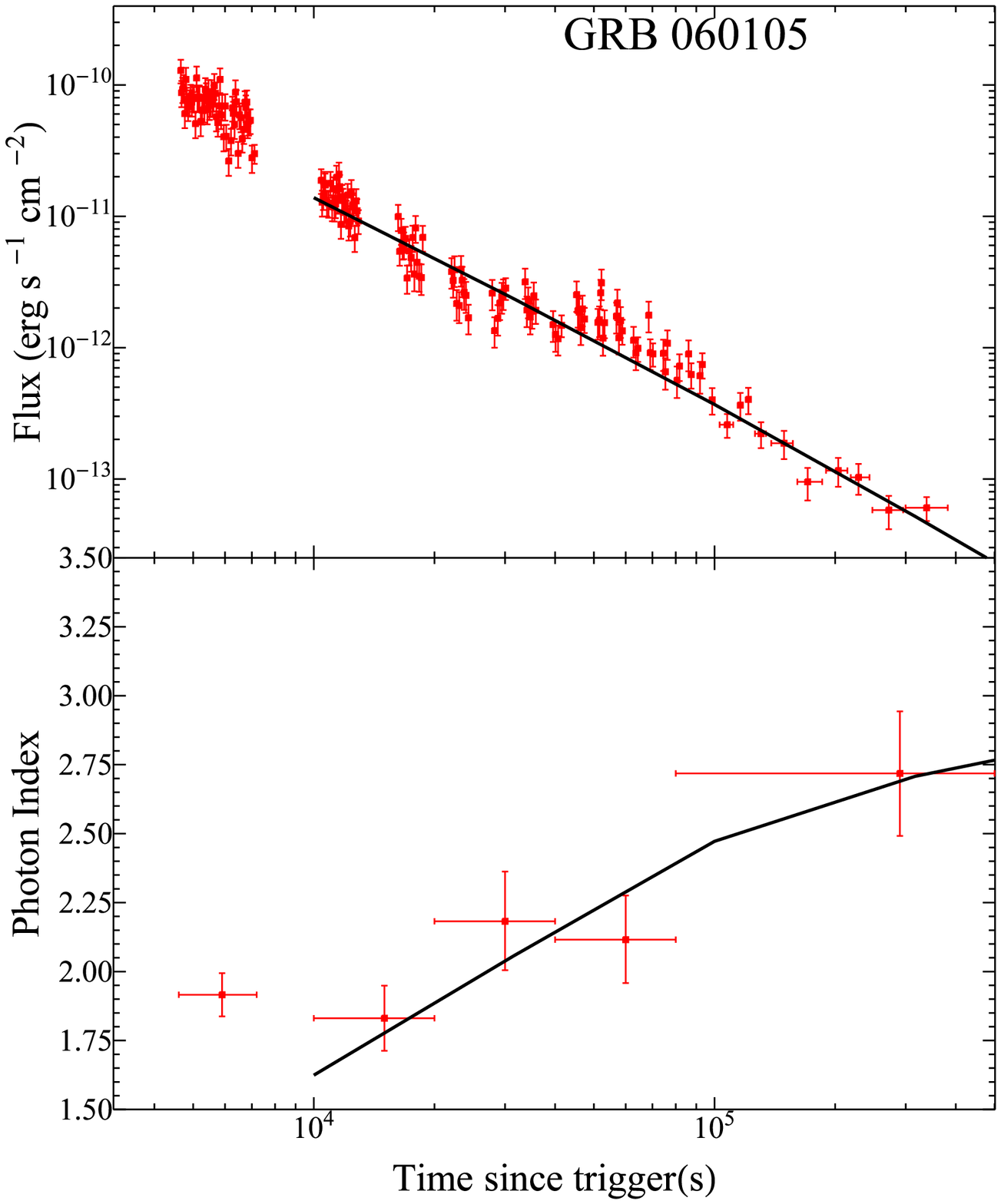}
\includegraphics[angle=0,scale=0.35]{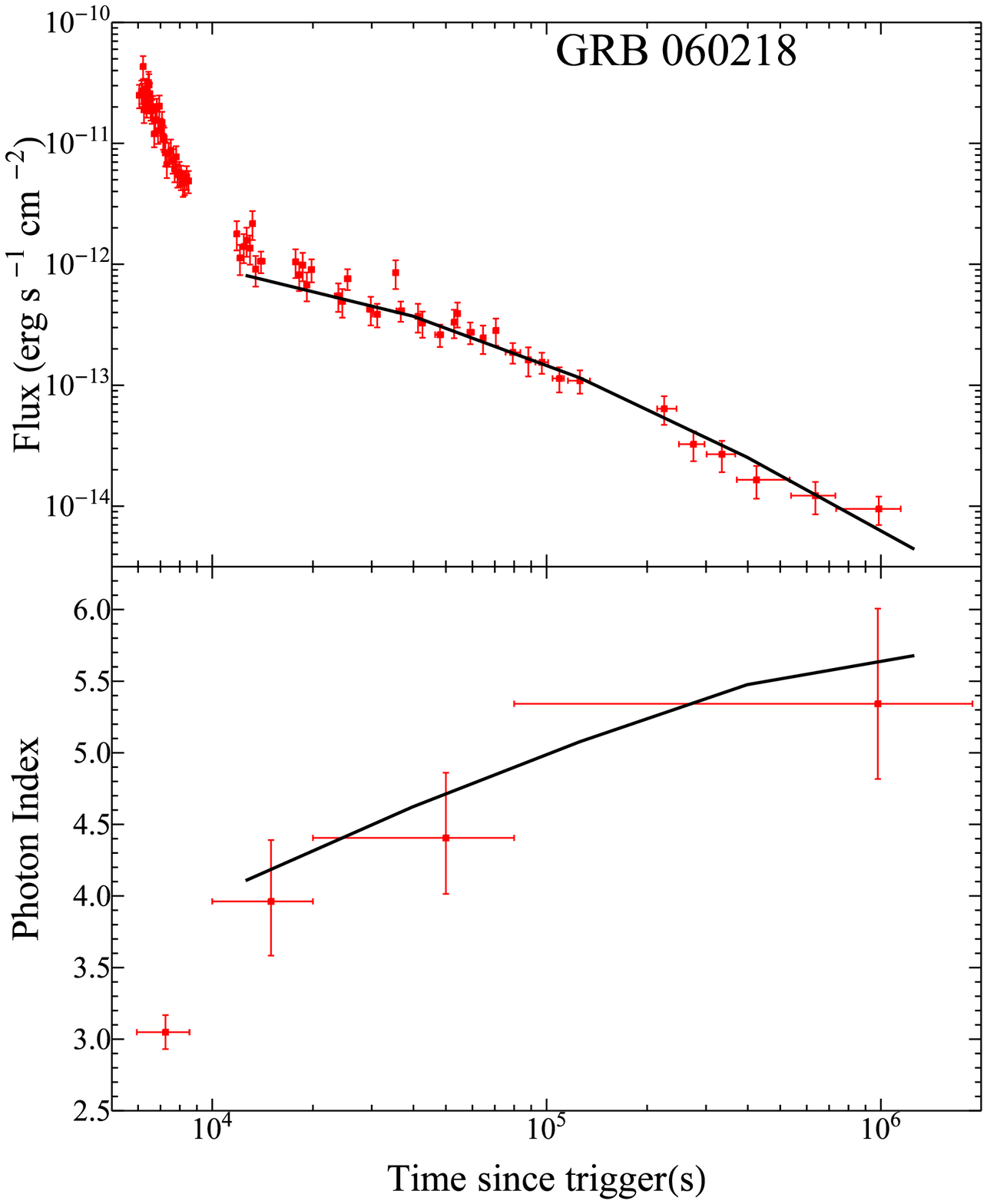}
\includegraphics[angle=0,scale=0.35]{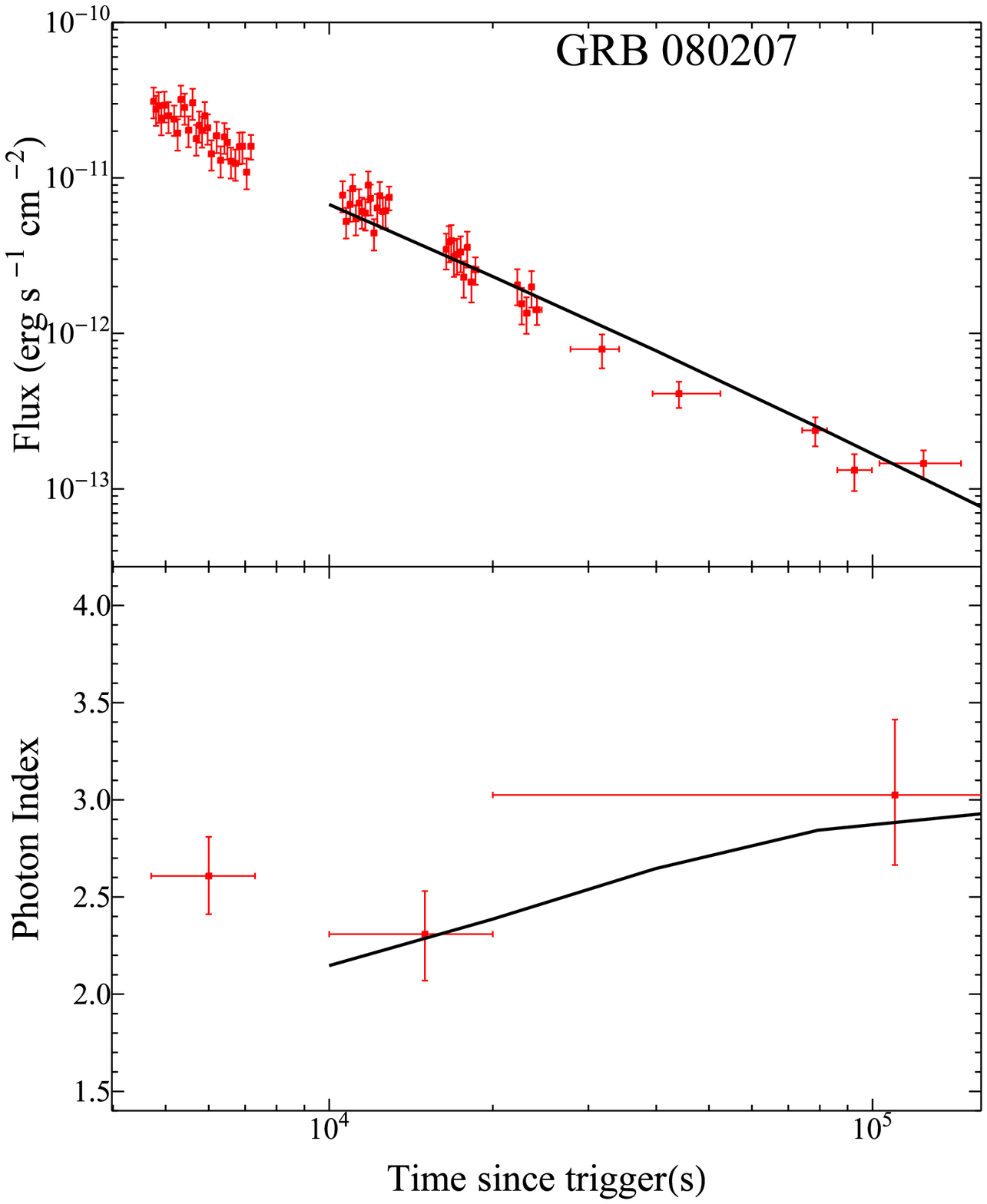}
\includegraphics[angle=0,scale=0.35]{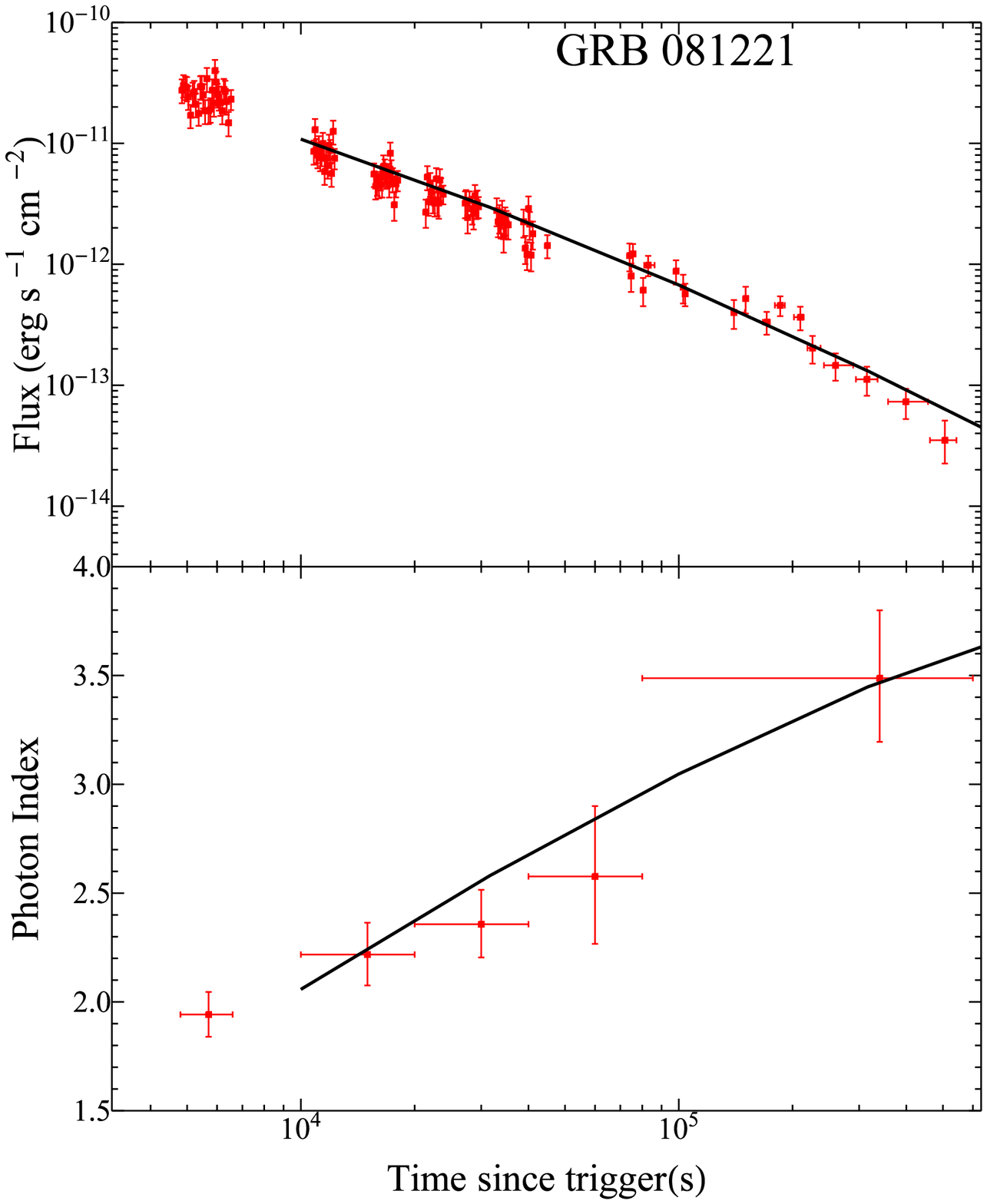}
\includegraphics[angle=0,scale=0.35]{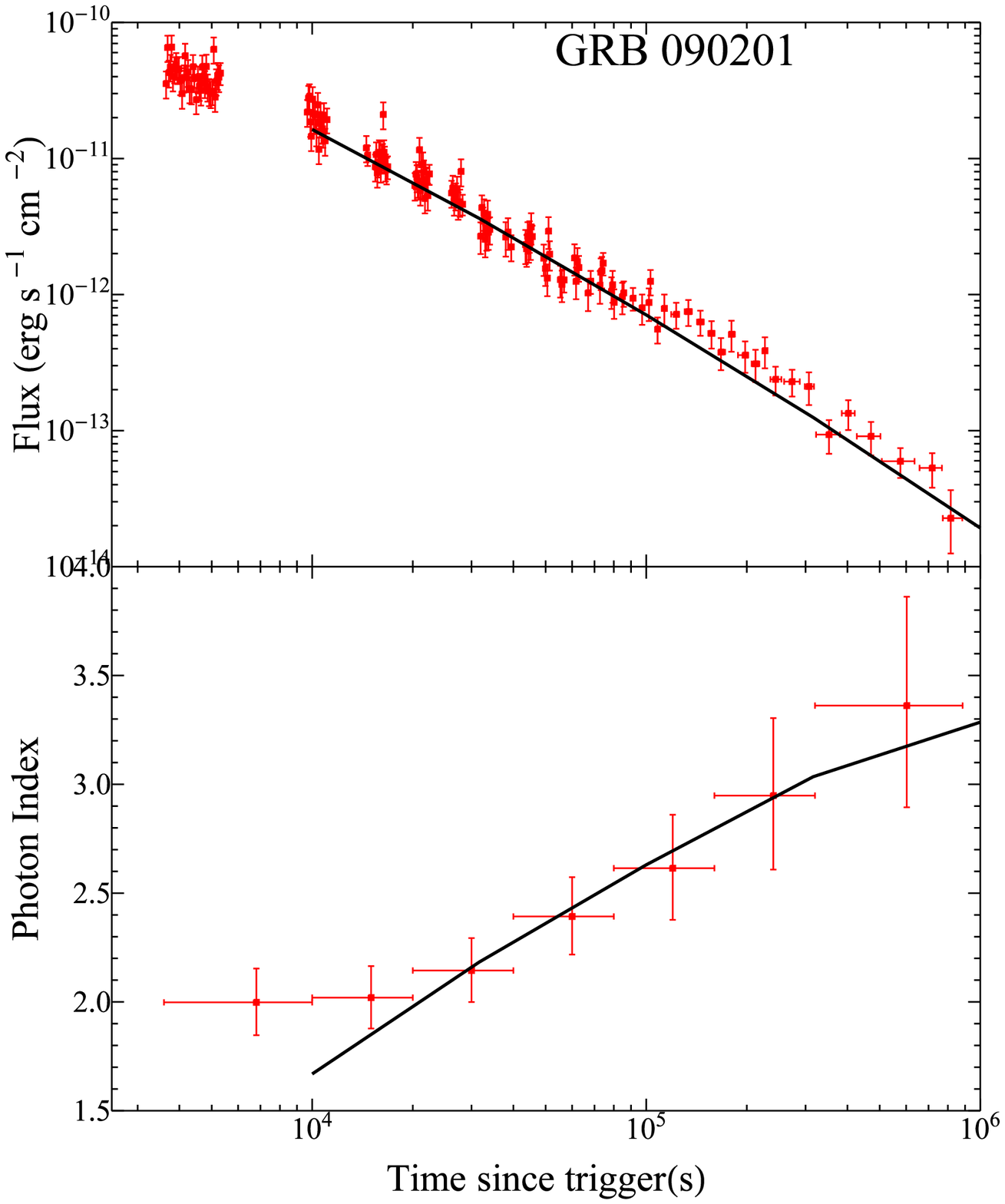}
\includegraphics[angle=0,scale=0.35]{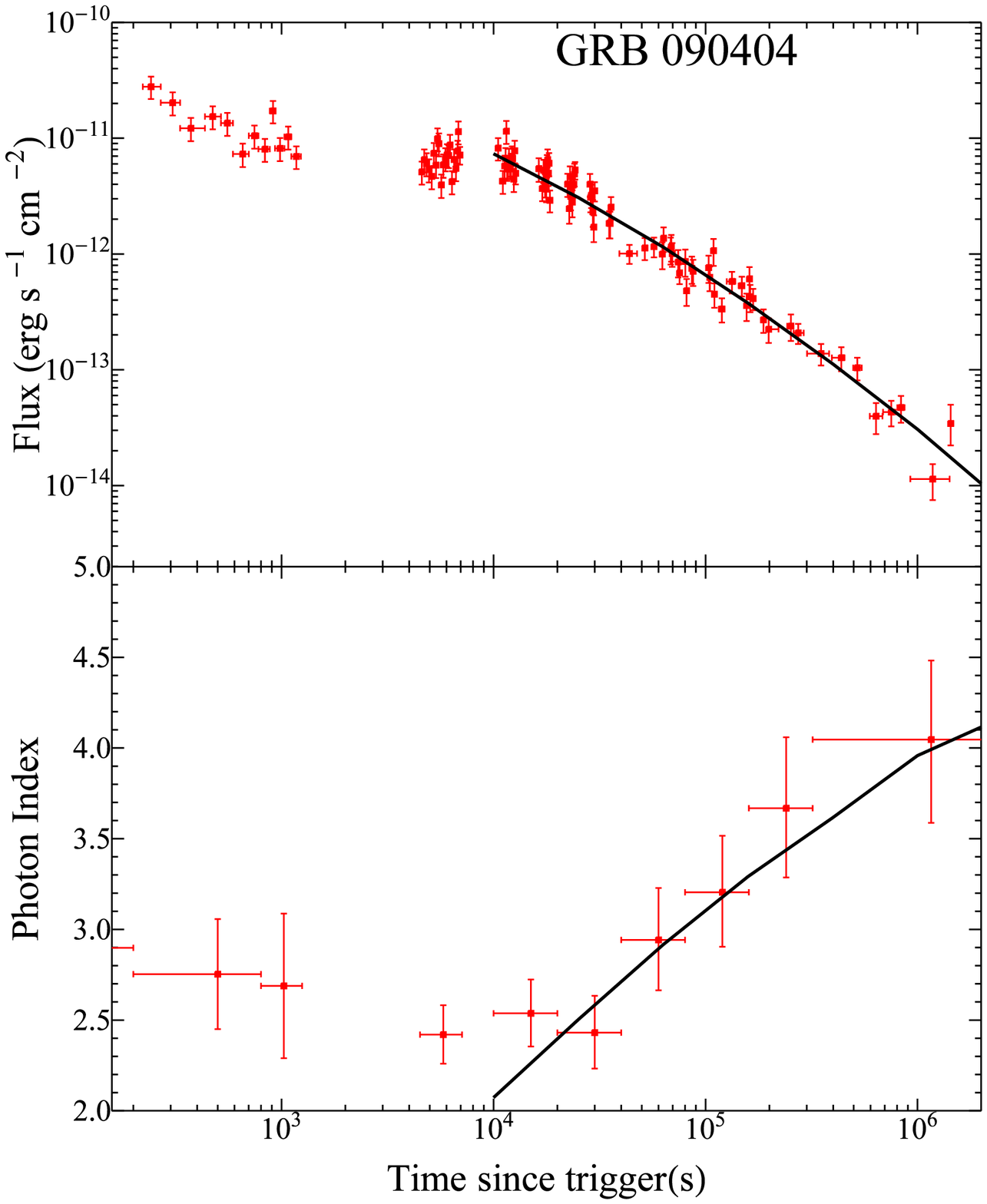}

\caption{X-ray light curves (upper panels) and evolution of spectral indices (lower panels) of eleven GRBs detected by {\it Swift}/XRT. The solid lines are the simultaneously best-fit results of dust scattering model.}
\hfill
\end{figure*}
\clearpage
\addtocounter{figure}{-1}

\begin{figure*}[htbp]
\centering
\includegraphics[angle=0,scale=0.35]{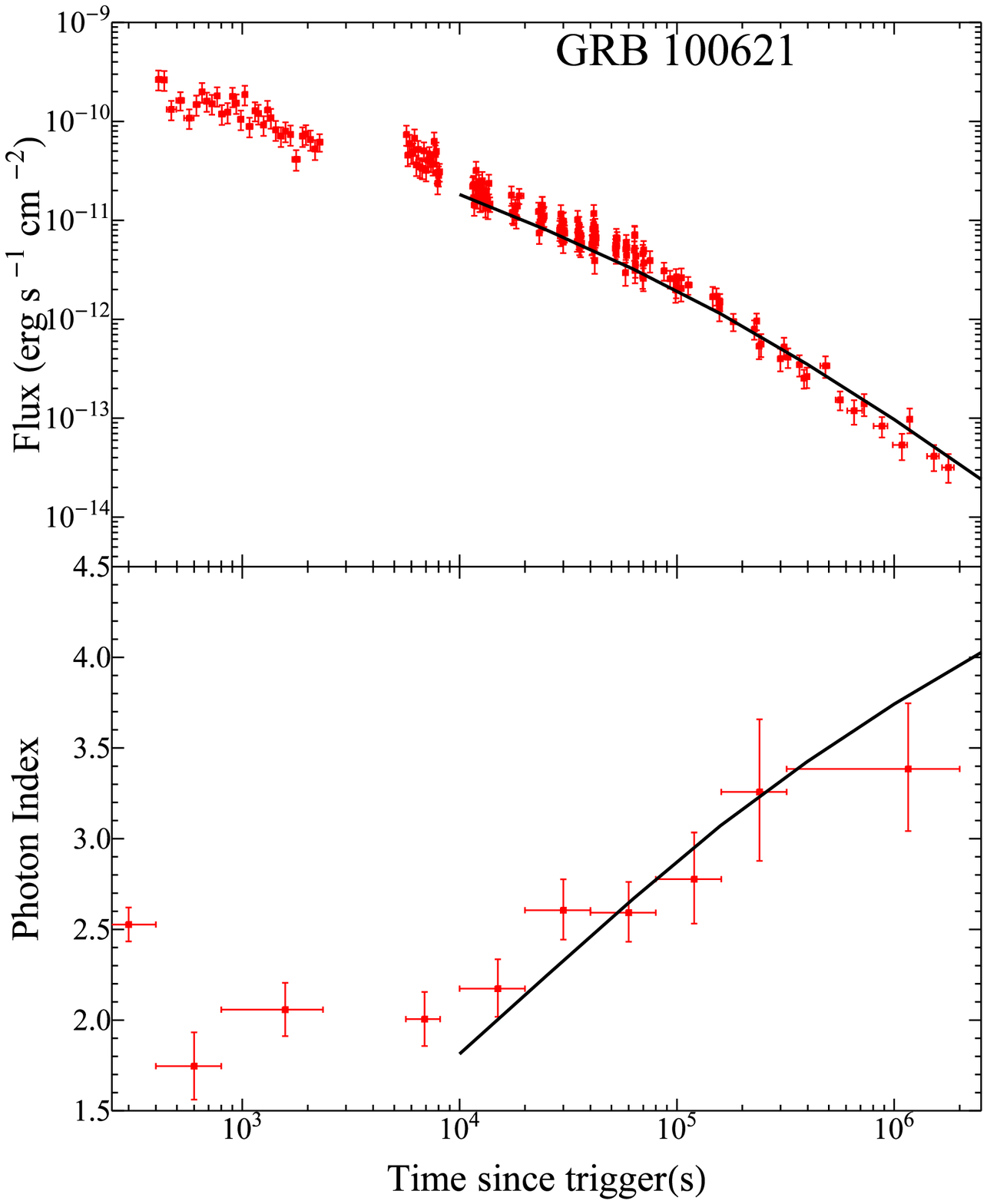}
\includegraphics[angle=0,scale=0.35]{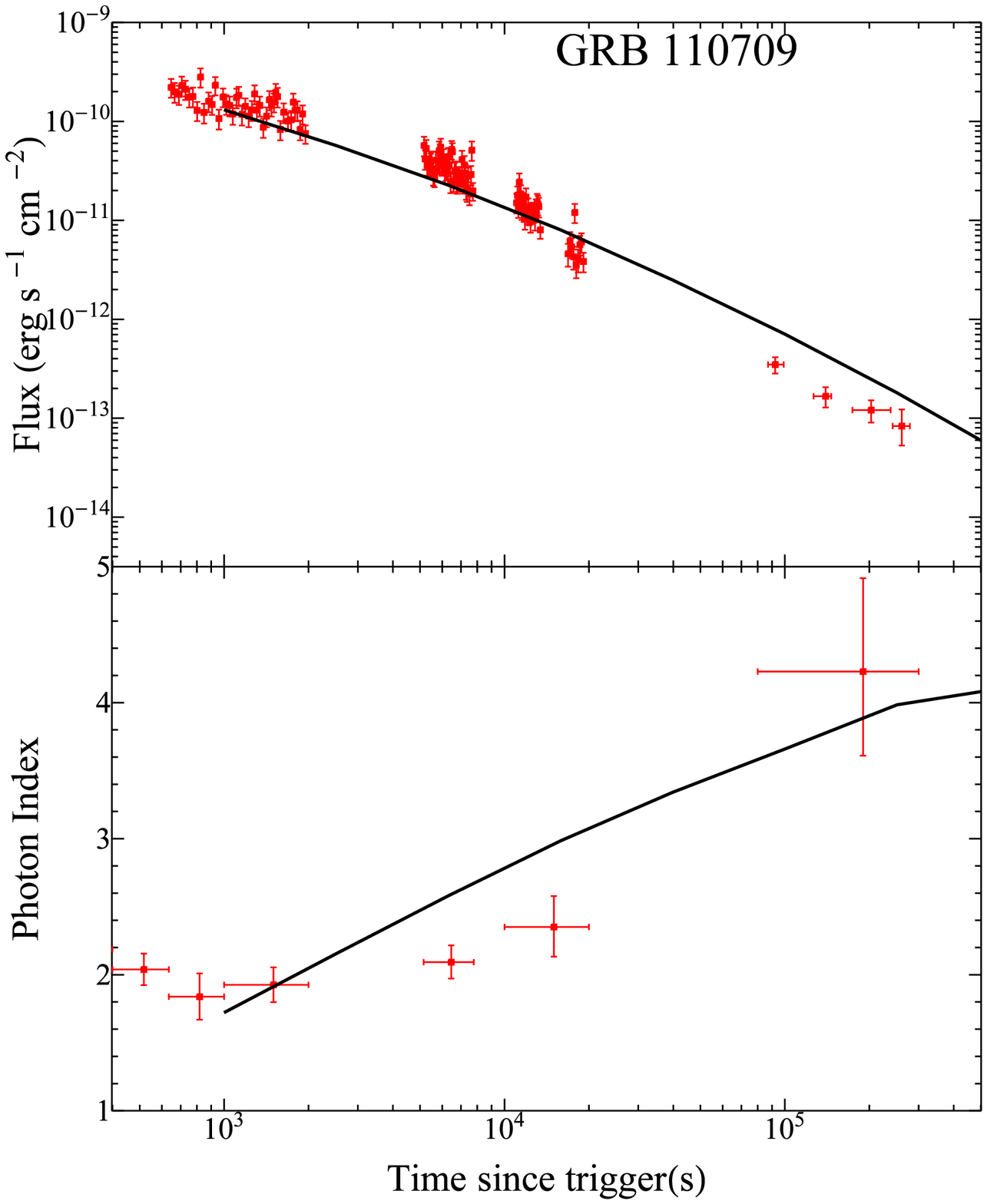}
\includegraphics[angle=0,scale=0.35]{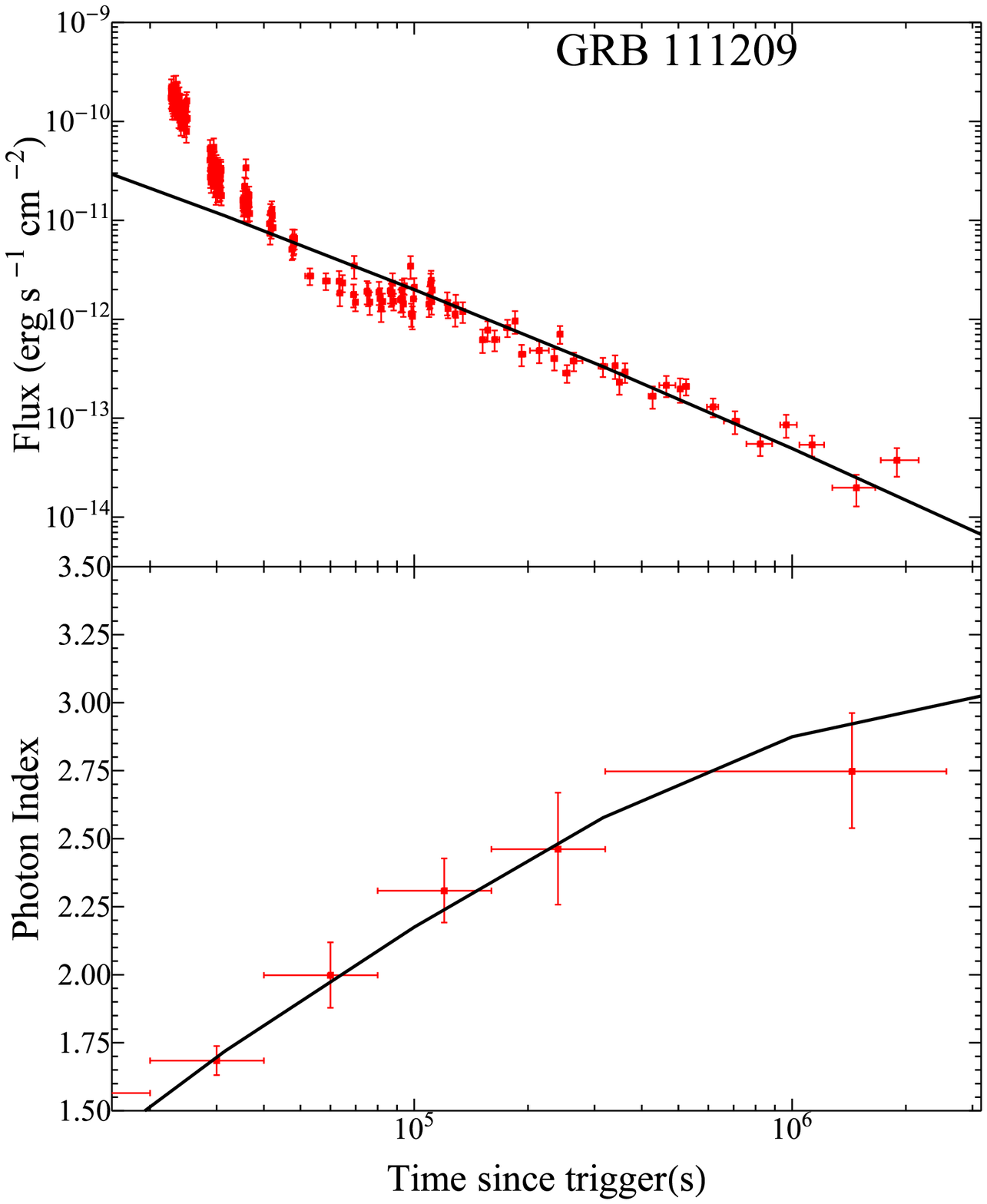}
\includegraphics[angle=0,scale=0.35]{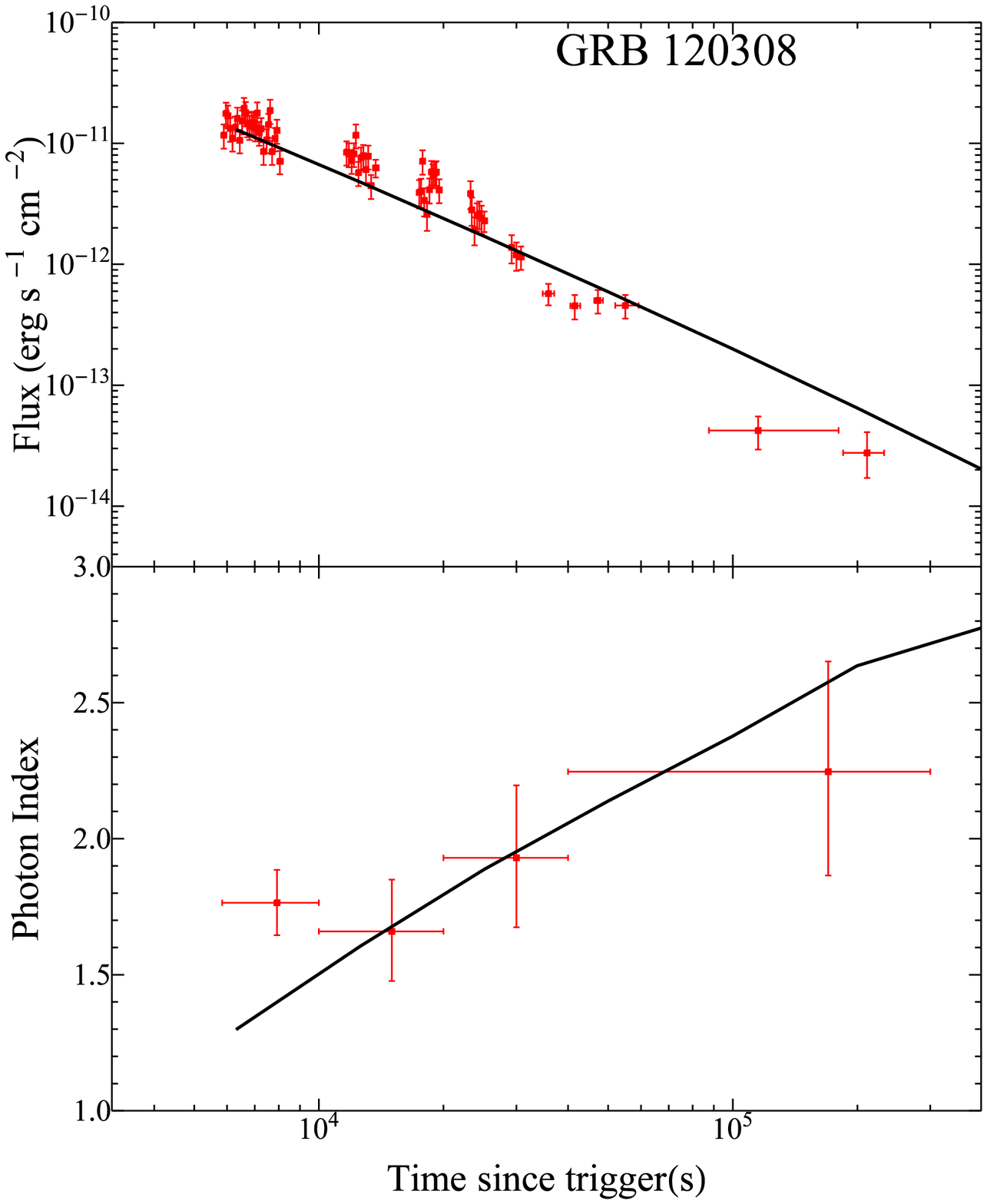}
\includegraphics[angle=0,scale=0.35]{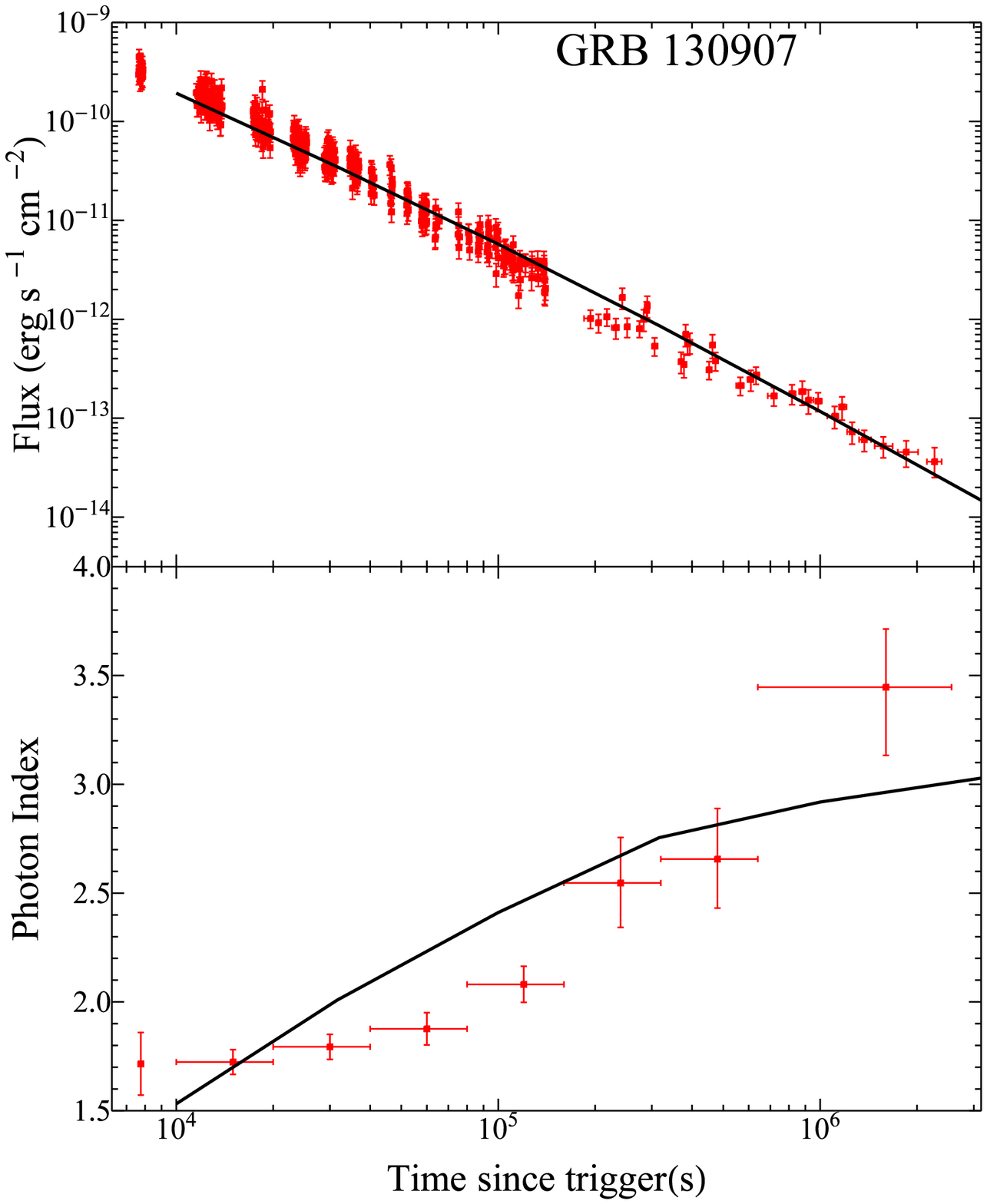}

\caption{Continued}
\hfill
\end{figure*}
\clearpage

\begin{figure*}[htbp]
\centering

\includegraphics[angle=0,scale=0.4]{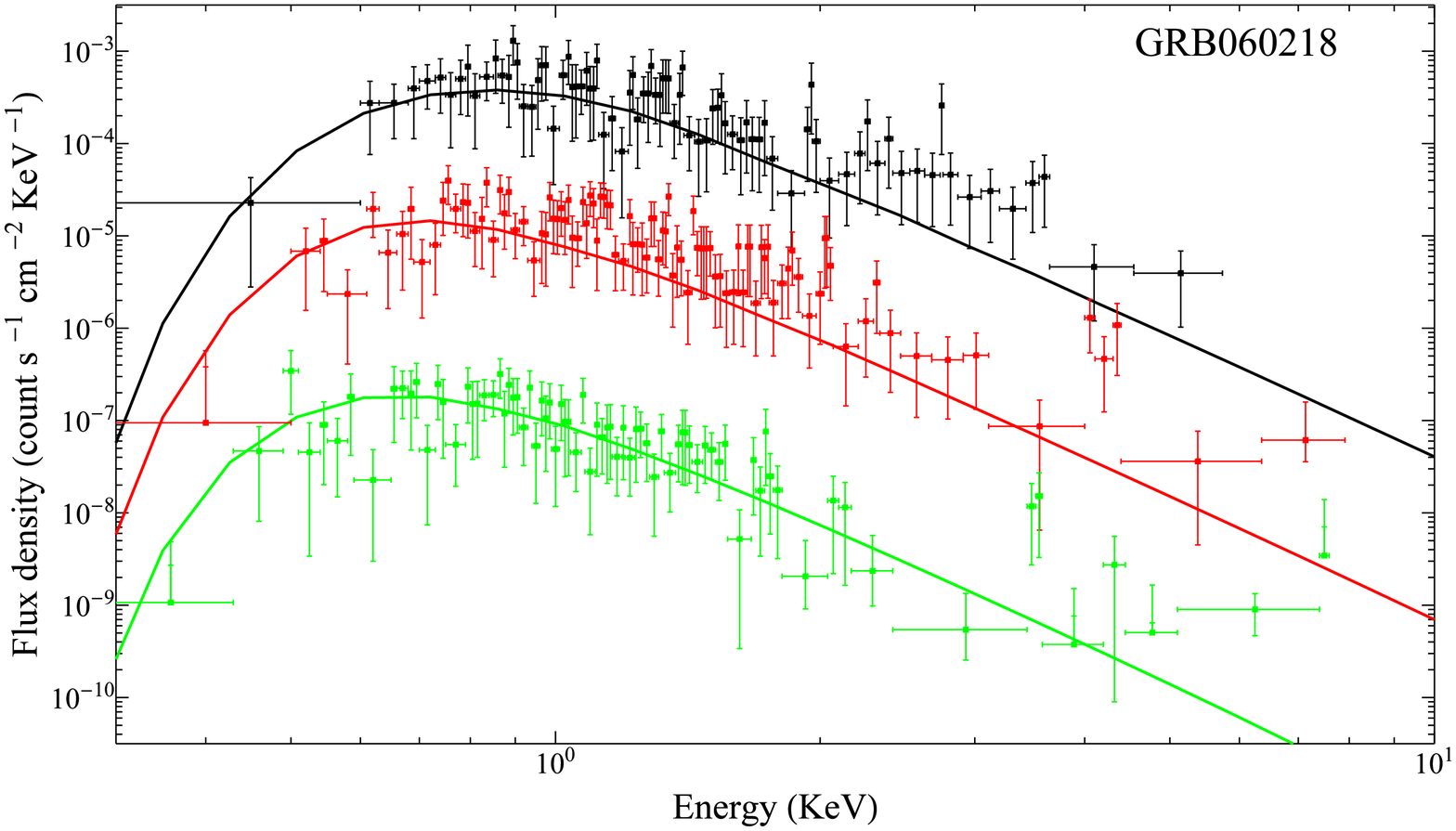}
\includegraphics[angle=0,scale=0.4]{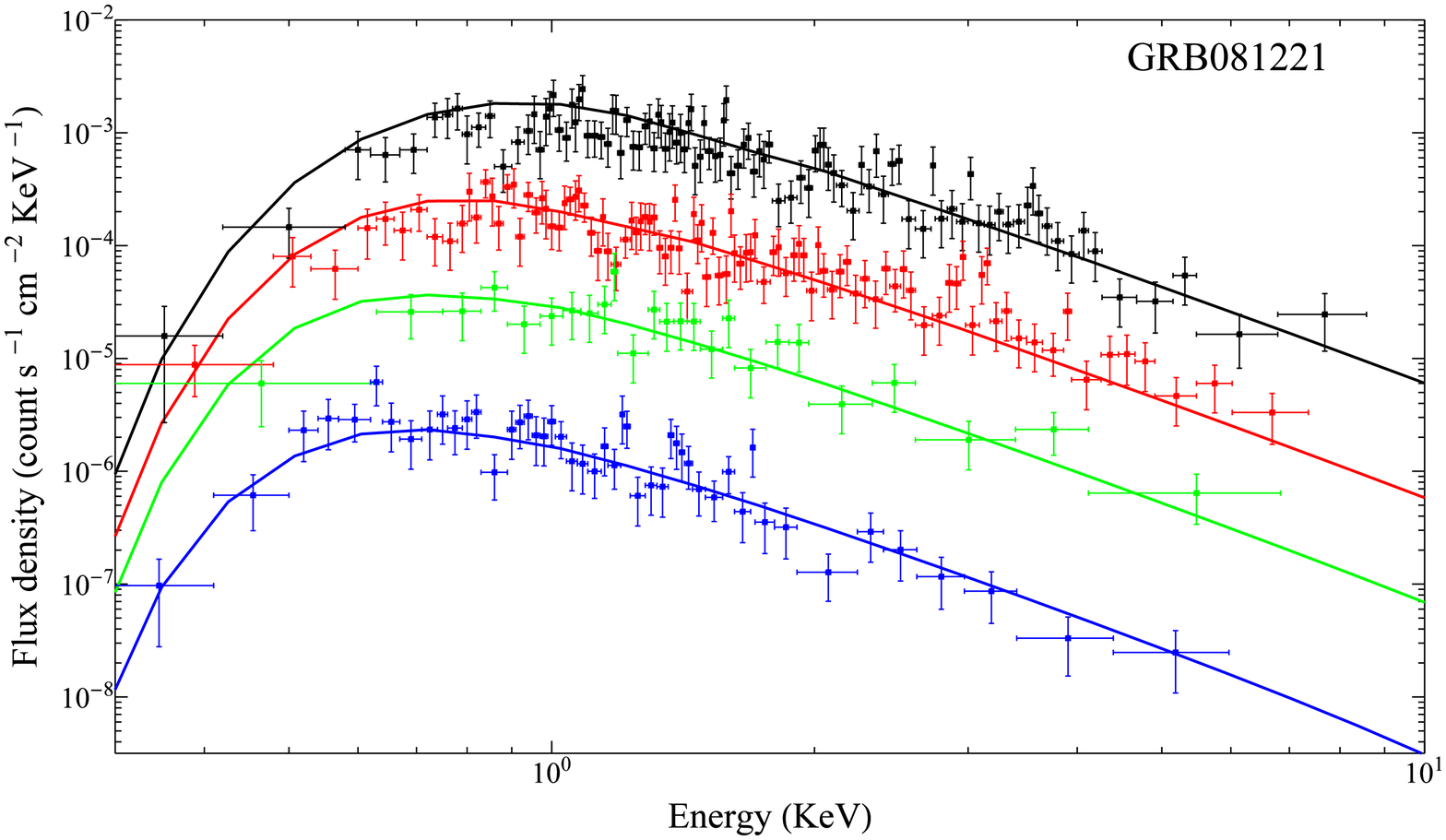}

\hfill
\caption{Time-resolved spectra of six GRBs within the different time intervals (from top to the bottom) listed in Table 3. The solid lines represent the simultaneously best-fit results of dust scattering model after correction for X-ray absorption as given by Eq.~(6). Different time intervals are indicated with different colors and artificial shifts have been added in the vertical scale to make each spectrum clearly separated from the others.}
\end{figure*}
\clearpage
\addtocounter{figure}{-1}

\begin{figure*}[htbp]
\centering

\includegraphics[angle=0,scale=0.4]{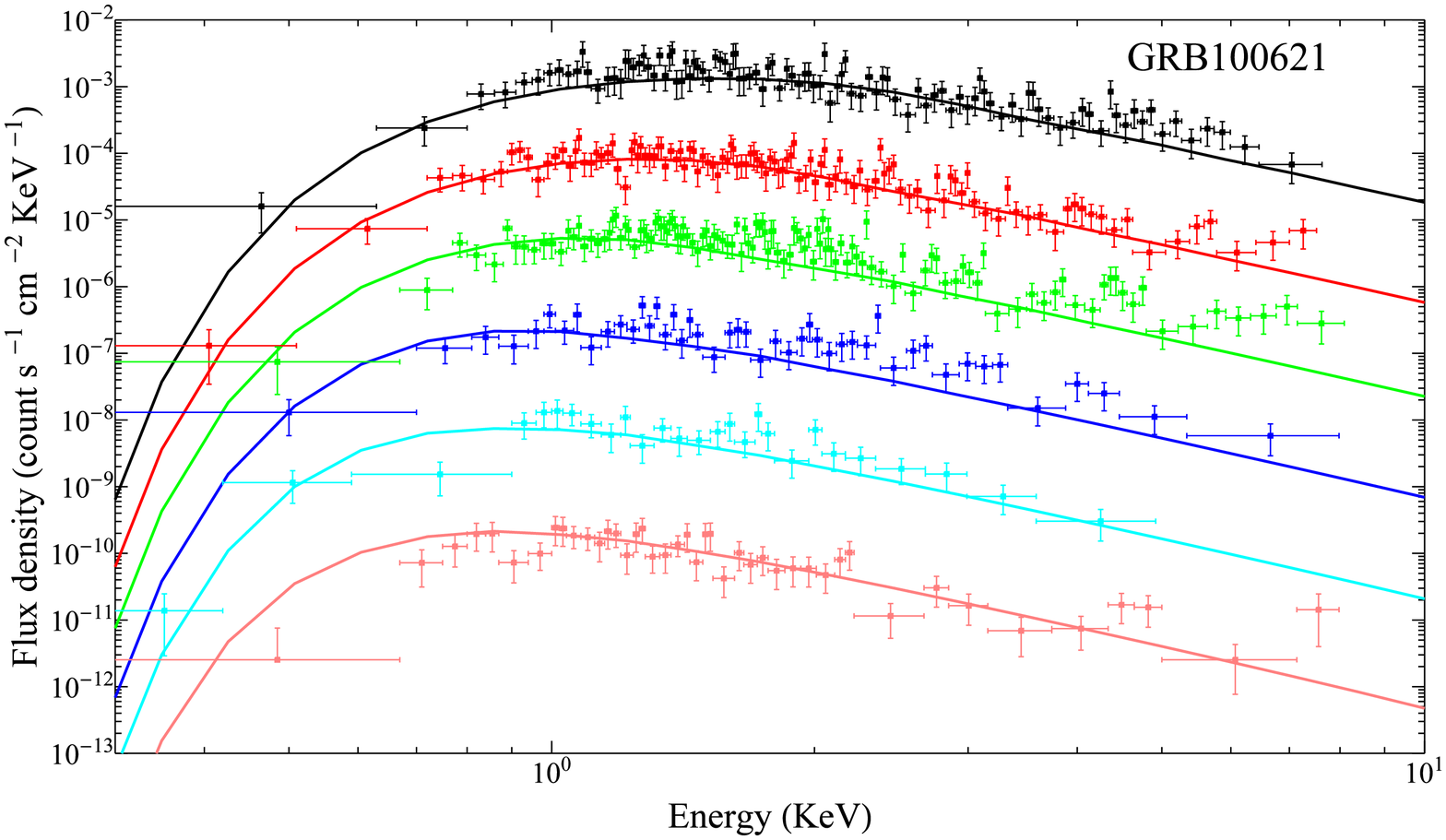}
\includegraphics[angle=0,scale=0.4]{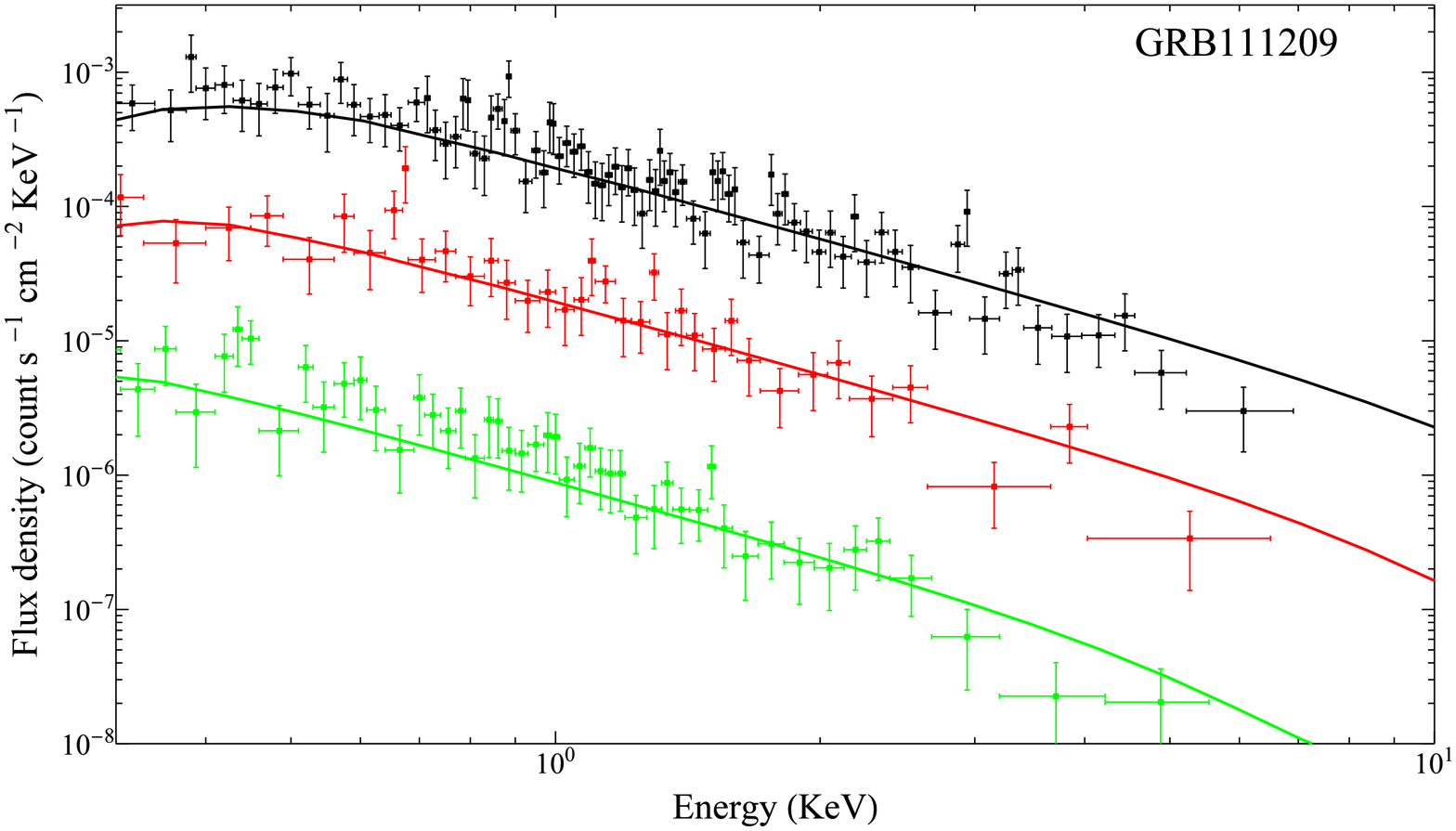}

\hfill
\caption{Continued}
\end{figure*}
\clearpage
\addtocounter{figure}{-1}

\begin{figure*}[htbp]
\centering

\includegraphics[angle=0,scale=0.4]{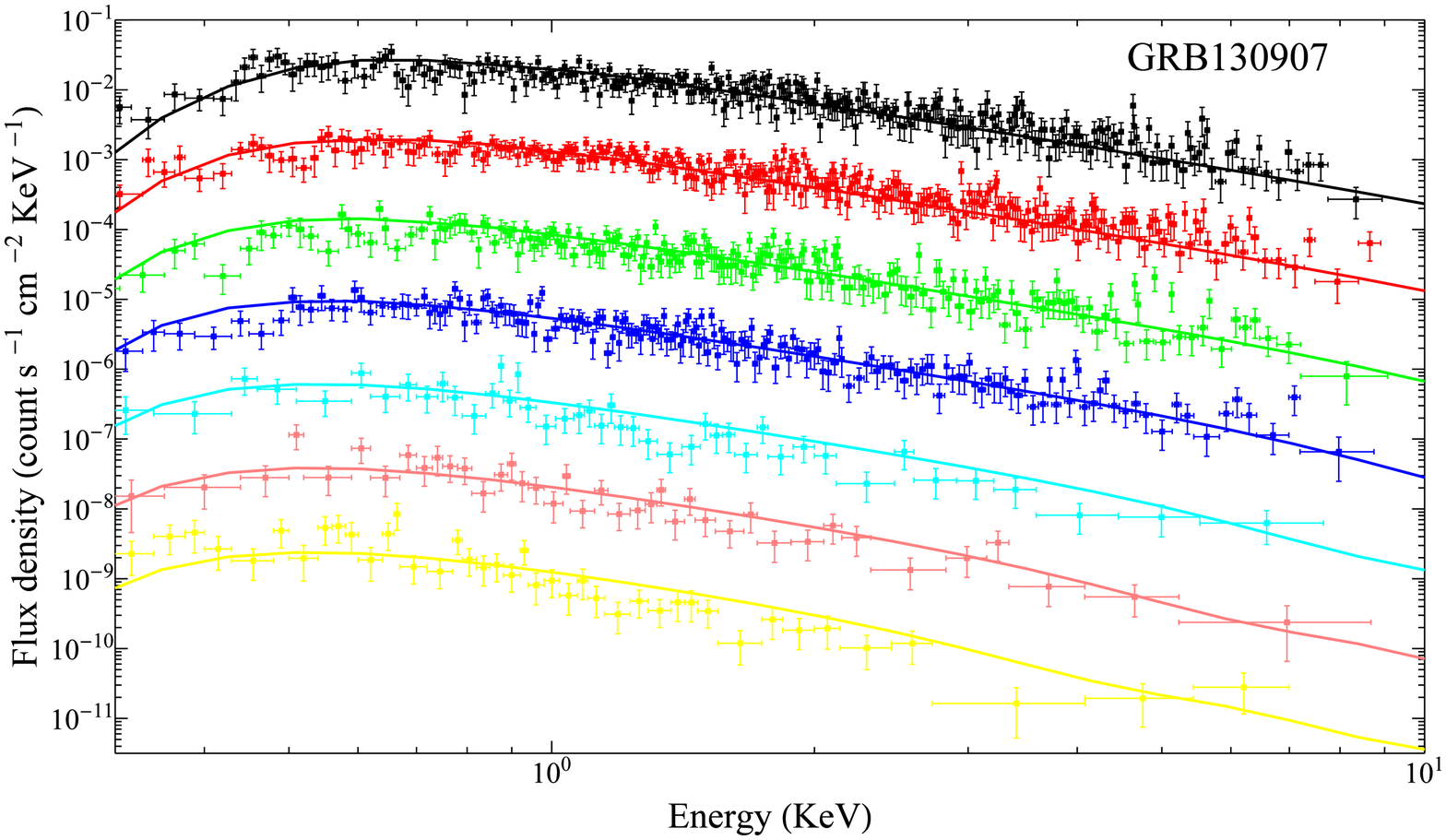}
\includegraphics[angle=0,scale=0.4]{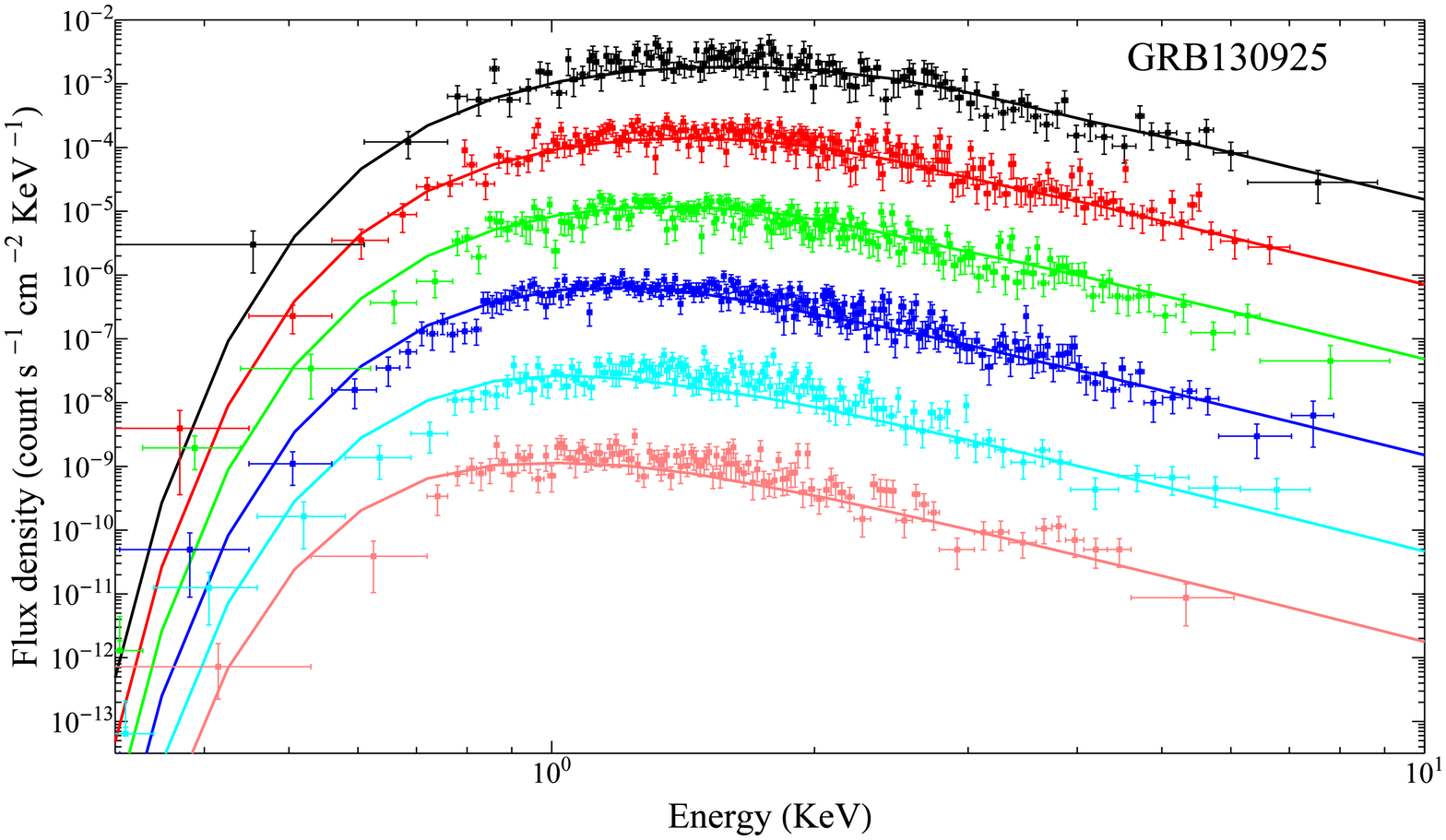}

\hfill
\caption{Continued}
\end{figure*}
\clearpage

\end{document}